\documentclass[fleqn,usenatbib]{mnras}

\usepackage{newtxtext,newtxmath}
\usepackage[T1]{fontenc}

\DeclareRobustCommand{\VAN}[3]{#2}
\let\VANthebibliography\thebibliography
\def\thebibliography{\DeclareRobustCommand{\VAN}[3]{##3}\VANthebibliography}

\usepackage{graphicx}	
\usepackage{amsmath}	
\usepackage[version=4]{mhchem}
\usepackage{gensymb}
\usepackage{chemfig}

\let\oldAA\AA
\renewcommand{\AA}{\text{\normalfont\oldAA}}

\title[3D brown dwarf chemical kinetics]{\centering{Dynamically coupled kinetic chemistry in brown dwarf atmospheres
\\ I. Performing global scale kinetic modelling}}

\author[Lee, Tan and Tsai]{
Elspeth K.H. Lee$^{1}$, Xianyu Tan$^{2,3}$ and Shang-Min Tsai$^{4}$ \\
$^{1}$Center for Space and Habitability, University of Bern, Gesellschaftsstrasse 6, CH-3012 Bern, Switzerland \\
$^{2}$Tsung-Dao Lee Institute, Shanghai Jiao Tong University, 520 Shengrong Road, Shanghai, People’s Republic of China \\
$^{3}$School of Physics and Astronomy, Shanghai Jiao Tong University, 800 Dongchuan Road, Shanghai, People’s Republic of China \\
$^{4}$Department of Earth Sciences, University of California, 900 University Ave, Riverside CA 92521, California, US
}

\date{Accepted XXX. Received YYY; in original form ZZZ}

\pubyear{2023}

\begin{document}
\label{firstpage}
\pagerange{\pageref{firstpage}--\pageref{lastpage}}
\maketitle

\begin{abstract}
The atmospheres of brown dwarfs have been long observed to exhibit a multitude of non-equilibrium chemical signatures and spectral variability across the L, T and Y spectral types.
We aim to investigate the link between the large-scale 3D atmospheric dynamics and time-dependent chemistry in the brown dwarf regime, and to assess its impact on spectral variability.
We couple the miniature kinetic chemistry module `mini-chem' to the Exo-FMS general circulation model (GCM). 
We then perform a series of idealised brown dwarf regime atmospheric models to investigate the dynamical 3D chemical structures produced by our simulations.
The GCM output is post-processed using a 3D radiative-transfer model to investigate hemisphere-dependent spectral signatures and rotational variability.
Our results show the expected strong non-equilibrium chemical behaviour brought on by vertical mixing as well as global spacial variations due to zonal flows.
Chemical species are generally globally homogenised, showing variations of $\pm$10\% or less, dependent on pressure level, and follow the dynamical structures present in the atmosphere.
However, we find localised storm regions and eddies can show higher contrasts, up to $\pm$100\%, in mixing ratio compared to the background global mean.
This initial study represents another step in understanding the connection between three-dimensional atmospheric flows in brown dwarfs and their rich chemical inventories.
\end{abstract}




\section{Introduction}

The study of chemical non-equilibrium processes in brown dwarf atmospheres has been an important factor in explaining their observed spectral features.
Several groups of molecules and their non-equilibrium properties have been examined in detail in the literature, in particular, \ce{CH4}-CO, \ce{NH3}-\ce{N2}-HCN and \ce{CO2}. 
Observed spectral feature strengths of each species group vary with the effective temperature and gravity of the brown dwarfs.

L dwarfs are within the effective temperature range of T$_{\rm eff}$ $\approx$ 2200-1400 K.
Generally, L dwarfs are characterised by their strong optical wavelength absorbers in their atmosphere such as Na, K as well as hydrides such as CaH and FeH, although the dominant absorbing species varies greatly with spectral subtype.
The strong TiO and VO bands seen in M dwarf spectra are weakened substantially in L dwarf spectra.
The most prominent infrared molecular features are \ce{H2O} and CO, found across the whole spectral type.

T dwarfs are within the effective temperature range of T$_{\rm eff}$ $\approx$ 1400-500 K, characterised spectroscopically with the onset of stronger \ce{CH4} absorption features \citep[e.g.][]{Burgasser2002}.
The early-T objects tend to better fit with cloudless or very thin cloud deck atmospheric models \citep[e.g.][]{Marley1996, Burgasser2022b}, suggesting the refractory clouds found in the L dwarf regime have been removed from the upper atmosphere.
However, models with minor clouds can also explain the spectrum of these objects well, in particular the later-T objects \citep[e.g.][]{Morley2012}.
The condensation of refractory elements removes the strong optical absorbers such as TiO, VO and Fe from the photospheric regions.
In the infrared, stronger \ce{CH4} features appear and begin to dominate the spectrum compared to CO.
However, hints of CO can still appear in the spectra of late T-dwarfs, suggesting their atmospheres are in chemical non-equilibrium \citep[e.g.][]{Geballe2009}.
For example, the well-studied T brown dwarf Gliese 229B (T$_{\rm eff}$ $\approx$ 900 K) shows clear signals of CO absorption in the atmosphere \citep[e.g.][]{Noll1997,Oppenheimer1998,Saumon2000}.
\citet{Calamari2022} were able to constrain the amount of CO on Gliese 229B using a retrieval model, and find a T$_{\rm eff}$ of the brown dwarf around 100 K smaller than the SED-derived value.
Recently, \citet{Miles2023} published a highly detailed spectrum of VHS 1256 b, which is a red, low gravity L-T transition type object, across the full infrared wavelength regime using a plethora of instruments onboard JWST.
They found evidence of disequilibrium chemistry from the strength of the \ce{CH4} spectral absorption features as well as a multitude of other species at non-equilibrium abundances.

Y dwarfs are brown dwarfs that have cooled to T$_{\rm eff}$ $\lesssim$ 500 K at this stage in their evolution. 
These are dominated by \ce{CH4} absorption and, due to their colder temperatures, have the possibility to contain water \citep[e.g.][]{Morley2018, Mang2022, Lacy2023} or KCl \citep[e.g.][]{Mang2022} clouds, which makes them interesting analogues to our own Solar System objects and more temperate climate objects.
Observational studies of late-T and Y brown dwarfs have also invoked chemical non-equilibrium processes in order to explain spectral feature strengths.
For example, in recent studies, \citet{Miles2020} perform M band spectroscopy on a selection of late-T and Y dwarfs (T$_{\rm eff}$ $\approx$ 750-250 K).
They find evidence of increased CO absorption across all their objects, not expected to be present when assuming chemical equilibrium.
They attempt to derive the eddy diffusion parameter, K$_{\rm zz}$, and CO quench levels from their observed spectral features.
\citet{Leggett2021} also showed the importance of taking non-equilibrium into account when fitting data to models. 
They find a modified deep adiabat with CO and \ce{NH3} in non-equilibrium can better fit the observed data trends in their Y-dwarf sample.

In the irradiated exoplanet regime, non-equilibrium chemistry is also expected, with photochemical processes also impacting the chemical structure of the atmosphere, with a recent example being the discovery and characterisation of \ce{SO2} on WASP-39b using JWST instruments \citep{Rustamkulov2023,Alderson2023,Tsai2022b}.
In 3D, \citet{Drummond2020} and \citet{Zamyatina2023} couple the \citet{Venot2019} thermochemical kinetic network to the UK Met Office Unified Model general circulation model (GCM) \citep{Mayne2014}. 
They focused on the canonical hot Jupiter exoplanets HD 209458b, HD 189733b, HAT-P-11b and WASP-17b.
They found all the planets showed vertical and zonal/meridional quenching with strong homogenisation of chemical species across the globe. 
Their HD 189733b simulation exhibited the most observable transport-induced quenching behaviour.
These 3D studies on hot Jupiter atmospheres are highly complementary to this study on brown dwarf atmospheres.

Overall, the major theme of the above and many more observational studies of brown dwarfs in the literature is the invocation of non-equilibrium chemical processes as an explanation for their observed spectral properties, in particular the minor species spectral features in the infrared.
In this study, we explore the 3D non-equilibrium chemistry of brown dwarf atmospheres by performing idealised simulations which couple a miniature kinetic chemistry module, mini-chem, to the Exo-FMS 3D GCM.
We perform five simulations across the effective temperature range: T$_{\rm eff}$ = 500 K, 750 K, 1000 K, 1250 K and 1500 K.

In Section \ref{sec:BDchem} we summarise approaches to modelling brown dwarf atmospheric chemistry. 
In Section \ref{sec:MC} and \ref{sec:GCM} we summarise the mini-chem kinetic chemistry module and the Exo-FMS GCM set-up for the brown dwarf simulations.
In Section \ref{sec:3Ddyn} we present the dynamical properties of each
simulations.
In Section \ref{sec:3Dchem} we present the 3D chemical properties and spatial distribution of the coupled model from simulation.
In Section \ref{sec:3Dspec} we present spectra of our simulation outputs, examining the equator and pole differences as well as rotational variation.
Section \ref{sec:disc} contains a discussion of our results.
Section \ref{sec:conc} contains the summary and conclusions of our study.

\section{Approaches to brown dwarf atmospheric chemistry}
\label{sec:BDchem}

The modelling of brown dwarf atmospheres has a large and expansive history \citep[e.g. see][for   comprehensive reviews]{Helling2014,Marley2015,Zhang2020}.
Most brown dwarf atmospheres to date have applied some version of chemical equilibrium assumption into their construction \citep[e.g.][]{Allard2001, Allard2011, Morley2012, Saumon2012, Morley2014, Phillips2020}.
Some models have included cloud models of various flavours, linked to equilibrium condensation for example, applying a `rainout' method where condensed elements are removed from the gas phase at pressures lower than their condensation pressure \citep[e.g.][]{Marley2002,Marley2021}.
Others have directly coupled cloud formation models such as the \citet{Ackerman2001} EddySed model \citep[e.g.][]{Stephens2009, Morley2012} and the kinetic cloud formation model DRIFT \citep[e.g.][]{Helling2008} in the case of the DRIFT-PHOENIX models \citep{Witte2009}.
These models also account for the removal of elements from the gas phase due to the cloud formation process.

A common practice in the literature to include the effects of non-equilibrium chemistry is to derive chemical timescales from a limiting (typically slow) reaction step that sets the quench level of a species.
This was first proposed for Jupiter by \citet{Prinn1977} to explain the detectable CO abundance in its atmosphere.
\citet{Fegley1996} then suggested similar processes to occur in brown dwarf atmospheres.
This technique or similar timescale analysis has been applied in several studies to estimate departures from chemical equilibrium in brown dwarf 1D atmospheric models \citep[e.g.][]{Hubeny2007,Phillips2020,Karalidi2021,Mukherjee2022,Lacy2023}.

\citet{Visscher2011} perform 1D chemical kinetics models of Gliese 229B and HD 189733b examining in detail the quench levels of CO and \ce{CH4} and the dependence on the strength of K$_{\rm zz}$ on the mixing ratio of the quenched species in the mid-upper atmosphere.
\citet{Zahnle2014} use a full 1D chemical kinetics model to explore a range of atmospheric parameters, from T$_{\rm eff}$ = 500-1100 K, K$_{\rm zz}$ = 10$^{4}$-10$^{11}$ cm$^{2}$ s$^{-1}$ and $\log$ g = 3-5 cm s$^{-2}$.
They focus on the finding quench points of key molecules such as \ce{CH4}, \ce{CO}, \ce{NH3}, \ce{N2}, HCN and \ce{CO2} and the accuracy and derivation of simple expressions for the quench points through analysis of chemical timescales.
These timescales have been extensively used for 1D atmospheric models.

3D approaches investigating chemical kinetics in this regime have been scarce in the literature.
\citet{Bordwell2018} perform 2D and 3D modelling of small-scale convection and mixing in brown dwarf and exoplanet regimes. 
They include a reactive tracer coupled to the flow to investigate the connection between parameterisations of 1D diffusion approximations (through the use of the K$_{\rm zz}$ eddy diffusion parameter) and the 2D/3D small scale flows in the convective zone of the atmosphere.

\section{mini-chem}
\label{sec:MC}

Mini-chem is an open source\footnote{\url{https://github.com/ELeeAstro/mini_chem}} miniature chemical kinetics solver for gas giant atmospheres, described in \citet{Tsai2022} and \citet{Lee2023}, developed as an offshoot of the VULCAN 1D chemical kinetics model \citep{Tsai2017, Tsai2021}.
In \citet{Lee2023} the thermochemistry of WASP-39b and HD 189733b were explored in 3D by coupling mini-chem to the Exo-FMS GCM model.
Mini-chem utilises `net forward rate' reaction tables, which greatly reduces the number of species and reactions (12 and 10 respectively for the C-H-N-O scheme) compared to other reduced kinetic schemes, for example, the \citet{Venot2019} network used in \citet{Drummond2020} contains 30 species with 181 reactions.
This comes at the cost of some accuracy and an assumption of a metallicity when constructing the net reaction tables \citep{Tsai2022}, but retains the classic chemical kinetic solver methodology.
Currently, the scheme only includes thermochemistry without photochemical reactions.

We include helium as a passive, inert, chemical tracer which is evolved with the flow and contributes to the total VMR, but does not alter any of the reacting species inside mini-chem. We integrate mini-chem every hour of simulation time, for a chemical timestep of $\tau_{\rm chem}$ = 3600 s. This allows a good compromise between computational efficiency and time-dependent accuracy. We use the seulex\footnote{\url{https://www.unige.ch/~hairer/software.html}} stiff ODE solver to integrate the network in time.

\section{GCM modelling}
\label{sec:GCM}

\begin{table*}
\centering
\caption{Adopted Exo-FMS GCM simulation parameters for the brown dwarf atmosphere simulations. We use a cubed-sphere resolution of C96 ($\approx$ 384 $\times$ 192 in longitude $\times$ latitude).}
\begin{tabular}{c c c l}  \hline \hline
 Symbol & Value  & Unit & Description \\ \hline
 T$_{\rm eff}$ & 500, 750, 1000 & K & Effective temperature \\
 & 1250, 1500 & & \\
 P$_{\rm 0}$ & 100 &  bar & Reference surface pressure \\
 P$_{\rm up}$ & 10$^{-4}$ &  bar & Upper boundary pressure \\
 P$_{\rm rcb}$ & 10 & bar & Radiative convective boundary \\
 c$_{\rm P}$ & 13000  &  J K$^{-1}$ kg$^{-1}$ & Specific heat capacity \\
 R & 3714 &  J K$^{-1}$ kg$^{-1}$  & Ideal gas constant \\
 $\kappa$ &  0.286 & -  & Adiabatic coefficient \\
 g$_{\rm bd}$ & 1000  & m s$^{-2}$ & Acceleration from gravity \\
 R$_{\rm bd}$ & 7.149 $\cdot$ 10$^{7}$  & m & Radius of brown dwarf\\
 $\Omega_{\rm bd}$ & 1.745 $\cdot$ 10$^{-4}$ & rad s$^{-1}$ & Rotation rate of brown dwarf \\
 M/H & 0 & - & $\log_{10}$ solar metallicity \\
 T$_{\rm amp}$ & 1.03 $\cdot$ 10$^{-5}$, 8.39 $\cdot$ 10$^{-5}$, 3.71 $\cdot$ 10$^{-4}$ & K s$^{-1}$ & Perturbation temperature amplitude \\
 & 1.2 $\cdot$ 10$^{-3}$, 3.0 $\cdot$ 10$^{-3}$ & & \\
 $\tau_{\rm storm}$ & 10$^{5}$ & s & Storm timescale \\
 $\Delta$ t$_{\rm hyd}$ & 60  & s & Hydrodynamic time-step \\
 $\Delta$ t$_{\rm rad}$  & 60 & s & Radiative time-step \\
 $\Delta$ t$_{\rm ch}$  & 3600 & s & Mini-chem time-step \\
 N$_{\rm v}$ & 60  & - & Vertical resolution \\
 d$_{\rm 4}$ & 0.16  & - & $\mathcal{O}$(4) divergence dampening coefficient \\
\hline
\end{tabular}
\label{tab:GCM_parameters}
\end{table*}

Overall, we follow a similar GCM set-up to the isolated brown dwarf MITgcm modelling efforts of \citet{Showman2019,Tan2021} and \citet{Tan2022}, adapted to the Exo-FMS GCM framework. 
Exo-FMS has been used to study terrestrial planets \citep{Hammond2017}, gas giant atmospheres \citep{Lee2021}, warm-Neptunes \citep{Innes2022} and irradiated brown dwarfs orbiting white dwarfs \citep{Lee2020}.
It is therefore well suited to perform the current suite of brown dwarf atmospheric simulations.
Table \ref{tab:GCM_parameters} shows the GCM parameters used for each simulation.

We use a two-stream, short characteristics method with B\'{e}zier interpolants \citep[e.g.][]{delaCruz2013} to calculate the vertical thermal radiation flux.
For the gas phase opacity we follow the same scheme as in \citet{Tan2022}, using the \citet{Freedman2014} Rosseland mean opacity fitting function with a minimum opacity of $\kappa_{\rm gas} = 10^{-3}$ m$^{2}$ kg$^{-1}$.
This grey limit is required to stabilise the scheme in the low optical depth limit, where the Rosseland mean is not as accurate at representing the atmospheric opacity and a Planck mean should be preferred \citep[e.g.][]{Heng2017}.
This method does not include scattering, however, significant thermal scattering is only expected in the presence of cloud particles, not applicable to the cloud-free models in this study.

We include the vertical convective transport of chemical tracers in-line with the dry convective adjustment scheme. 
This mixes chemical species in a way parallel to the temperature changes when the dry convection adjustment is triggered by homogenising the VMR of species between vertical layers that undergo adjustment.
As in the temperature adjustment, the scheme is performed until the vertical temperature-pressure profile is stable to convection.
This operates instantaneously, the same way as the temperature adjustment is performed in the convective adjustment scheme.
As a result, chemical tracers are highly efficiently mixed inside the adiabatic region as expected from smaller scale simulations in similar regimes \citep[e.g.][]{Freytag2010, Bordwell2018}.
We note the GCM model cannot resolve individual convective plumes which would be much smaller than the grid resolution used here. 

In addition, we also note that different approaches to tracer mixing in convective regions exist such as mixing length theory \citep[e.g. see][]{Marley2015}.
Using mixing length theory would allow tracers to mix through the calculation of an eddy diffusion coefficient, K$_{\rm zz}$, derived from the convective heat flux.
However, in this study, we retain the convective adjustment scheme as it is  simpler to implement inside the GCM framework.
Examining the differences between a mixing length theory approach and convective adjustment inside the GCM is left to future studies.

To induce a dynamical response in the atmosphere we follow the thermal perturbation forcing scheme used in previous brown dwarf dynamics studies \citep{Showman2019, lian2022, Tan2022}. 
This scheme aims to approximate the perturbations induced through convective plume motions from deeper in the atmosphere. The term `forcing' in the context of atmospheric dynamical modeling generally refers to the way the sources of thermodynamic heating and cooling or angular momentum are implemented to drive the atmospheric flow.
We follow directly the vertical perturbation form in \citet{Tan2022}, choosing a constant radiative-convective boundary pressure of 10 bar for each simulation and storm timescale, $\tau_{\rm storm}$, of 10$^{5}$ s.
The thermal perturbation amplitude is changed dependent on the effective temperature (Table \ref{tab:GCM_parameters}), which is calculated following the approximations found in \citet{Showman2019}.
Without this perturbation scheme, the atmosphere remains static in radiative-convective equilibrium without evolvement of the atmospheric wind structure.

We apply a deep `linear basal drag', implemented as a Rayleigh drag term as in \citet{Tan2021}, using a surface drag at the lower boundary (100 bar) which is linearly decreased in strength to a pressure of 50 bar.
This emulates a moderate deep drag from magnetic and other drag sources on the atmosphere \citep[e.g.][]{Beltz2022}.
Such schemes are commonly used in gas giant planet atmospheric simulations \citep[e.g.][]{Liu2013,Komacek2016,Carone2020}.
We chose a `middle ground' drag timescale of $\tau_{\rm drag}$ = 10$^{6}$ s to allow larger scale dynamical features to develop, compared to the more dampened dynamics exhibited by smaller drag timescales \citep{Tan2022}.
Other drag timescale values, 10$^{5}$ s and 10$^{7}$ s, were explored in \citet{Tan2021, Tan2022}, and its impact on the atmospheric dynamics was assessed.
Kinetic energy removed through this deep Rayleigh drag is returned to the atmosphere as localised increases in thermal energy.

{Rotation periods of brown dwarfs span from about 1 hour to nearly 20 hours according to spectroscopic and photometric observations \citep[e.g.][]{reiners2008,Metchev2015,tannock2021}. 
We chose a rotation period of 10 hours as a representative value in this work, following a middle ground value from the \citet{Tan2021} suite of brown dwarf models. 
A higher rotation rate typically results in smaller scale horizontal dynamical features, requiring a greater horizontal resolution in the GCM to adequately resolve. 
Our assumed moderate rotation period allows us to avoid using a higher GCM grid resolution than C96, required for accurate capture of smaller scale features that appear at such a rotation period while retaining strong rotationally driven flows.
Increasing the rotation rate and resolution of the GCM would require a large boost in available computational power to perform.}

With a drag timescale of 10$^{6}$ s, the atmosphere achieves statistical equilibrium after around 1000 days of simulation \citep{Tan2022}. 
All GCM models are therefore run for 1000 simulated days before coupling to mini-chem is included.
We assume chemical equilibrium as the VMR initial conditions for each species in the atmosphere starting at 1000 days, after which mini-chem is used to evolve the VMR composition of the atmosphere. 
During testing, it was found that the kinetics solver was too slow with integrating high-pressure, high-temperature deep atmospheric regions where the chemical timescales are very small, with the exception of the T$_{\rm eff}$ = 500 K and, surprisingly, the T$_{\rm eff}$ = 1500 K model. 
We therefore enforce chemical equilibrium for pressure levels greater than $\approx$20 bar in the 750, 1000 and 1250 K models to ensure a smooth VMR profile and expedient computational completion time.

We run the coupled scheme until 2100 days for the T$_{\rm eff}$ = 500 K, 750 K and 1000 K simulations and 1900 and 1400 days for the T$_{\rm eff}$ = 1250 K and 1500 K models respectively.
We find that the T$_{\rm eff}$ = 1250 K and 1500 K models converge faster compared to the cooler models due to being forced with a larger temperature perturbation amplitude.

\section{3D dynamical structures}
\label{sec:3Ddyn}

In this section, we present the dynamical structures of the five brown dwarf GCM simulations.

\subsection{Zonal mean zonal velocity}

\begin{figure*} 
   \centering
   \includegraphics[width=0.49\textwidth]{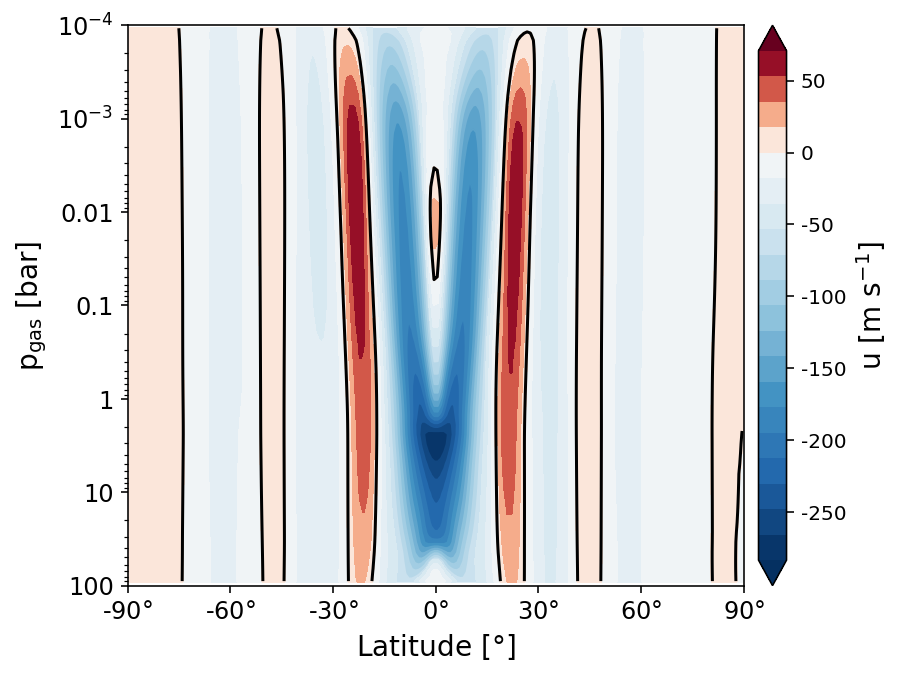}
   \includegraphics[width=0.49\textwidth]{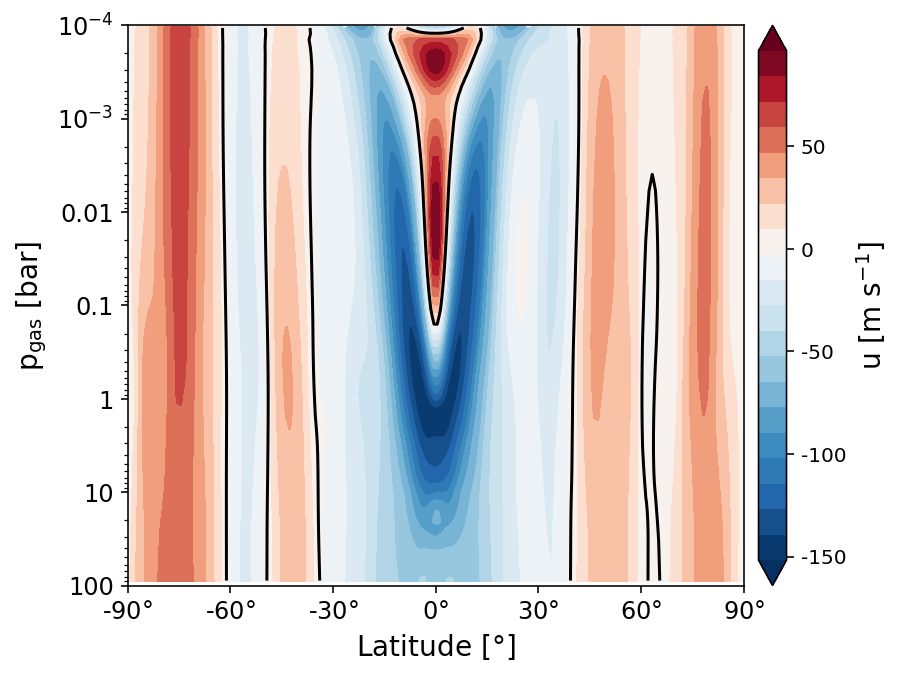}
   \includegraphics[width=0.49\textwidth]{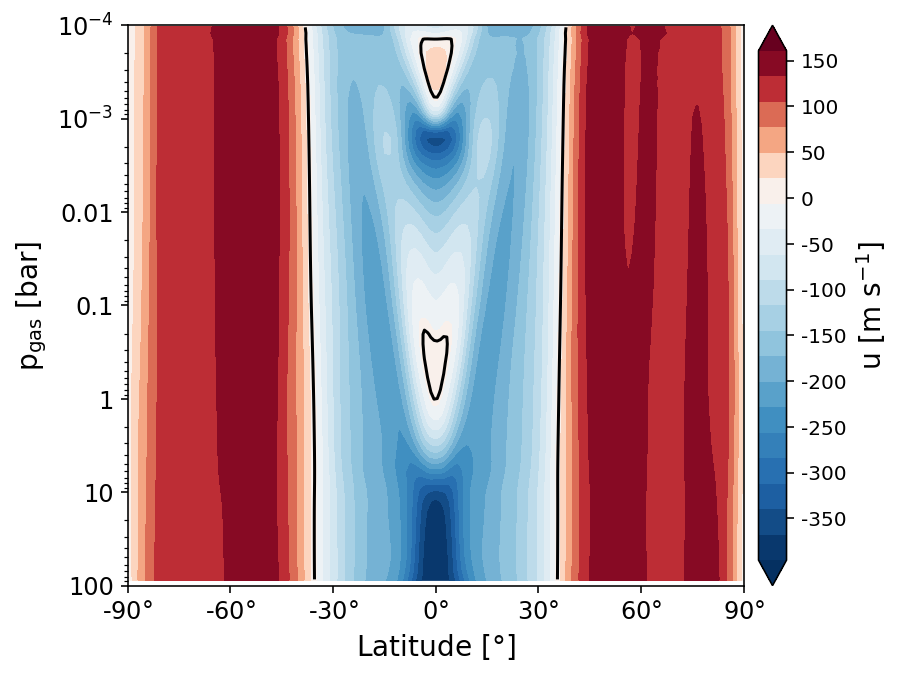}
   \includegraphics[width=0.49\textwidth]{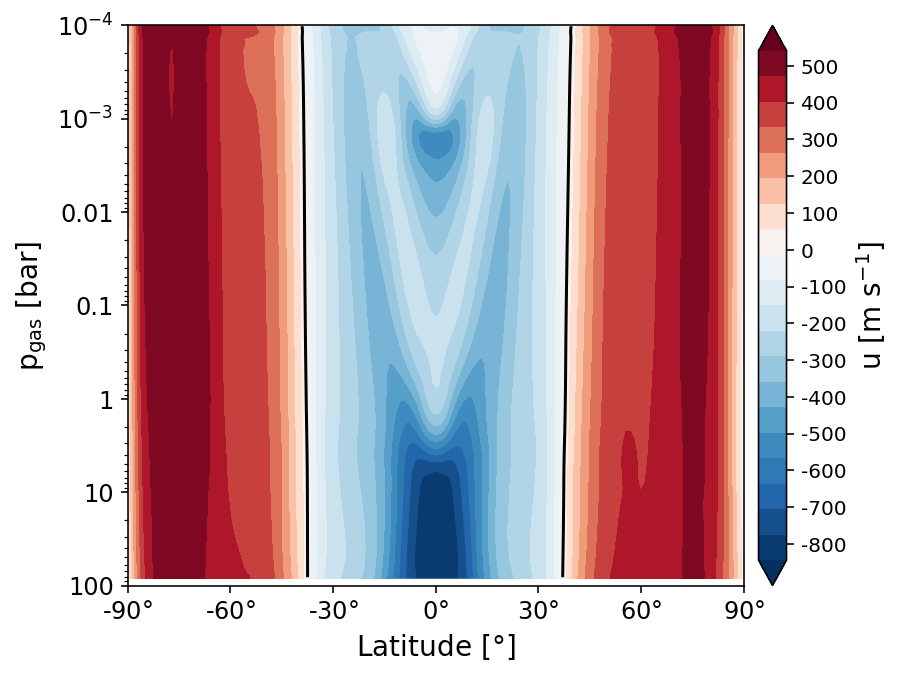}
   \includegraphics[width=0.49\textwidth]{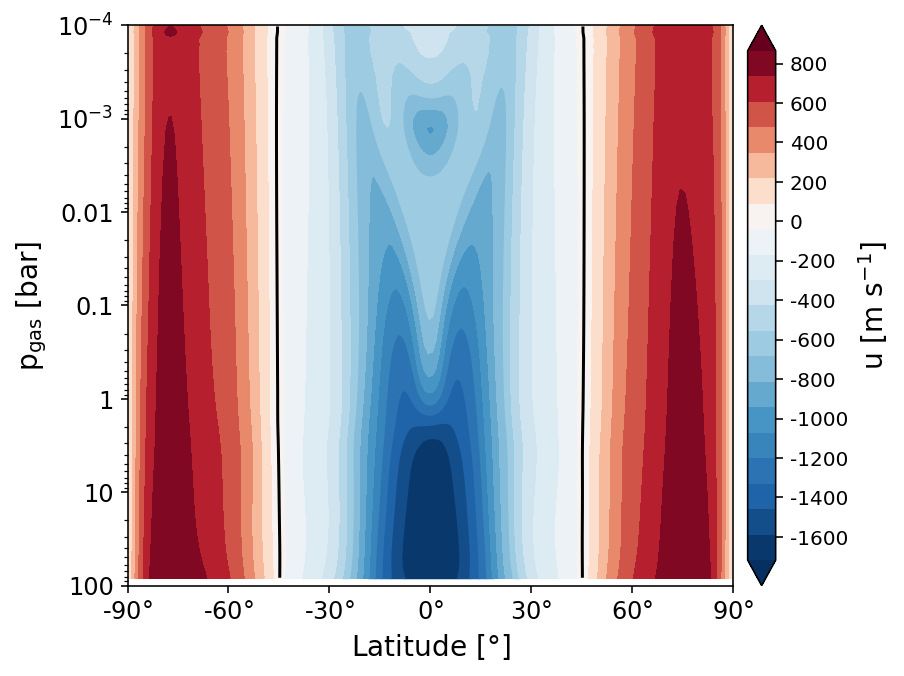}
   \caption{Zonal mean zonal velocity of each brown dwarf simulation averaged from the last 100 days of simulation. 
   T$_{\rm eff}$ = 500 K (top left), 750 K (top right), 1000 K (middle left), 1250 K (middle right) and 1500 K (bottom).
   The strongest forced models (1000 K, 1250 K, 1500 K) show large scale jet formation patterns from the equator, with one counter rotating jet at equatorial regions, with flanking jets at higher latitudes.
   The cooler models (500 K, 750 K) show more evidence of multiple jet structure formation.}
   \label{fig:zonal}
\end{figure*}

In Figure \ref{fig:zonal} we show the zonal mean zonal plots of each of the five brown dwarf GCMs. 
This gives an idea of the jet structures and patterns present in each simulation.
The T$_{\rm eff}$ = 500 K simulation, shows evidence for a multiple alternating pro- and retro-grade jet structure, typical of a fast rotator and reminiscent of Jupiter's multiple zonal bands.
The other simulations show a wide equatorial jet structure with flanking jets at latitude.
This is in line with the stronger temperature perturbation amplitude zonal mean results from \citet{Tan2021} and \citet{Tan2022}, which show similar structures in our chosen parameter regime.

\subsection{Outgoing longwave radiation}

\begin{figure*} 
   \centering
   \includegraphics[width=0.49\textwidth]{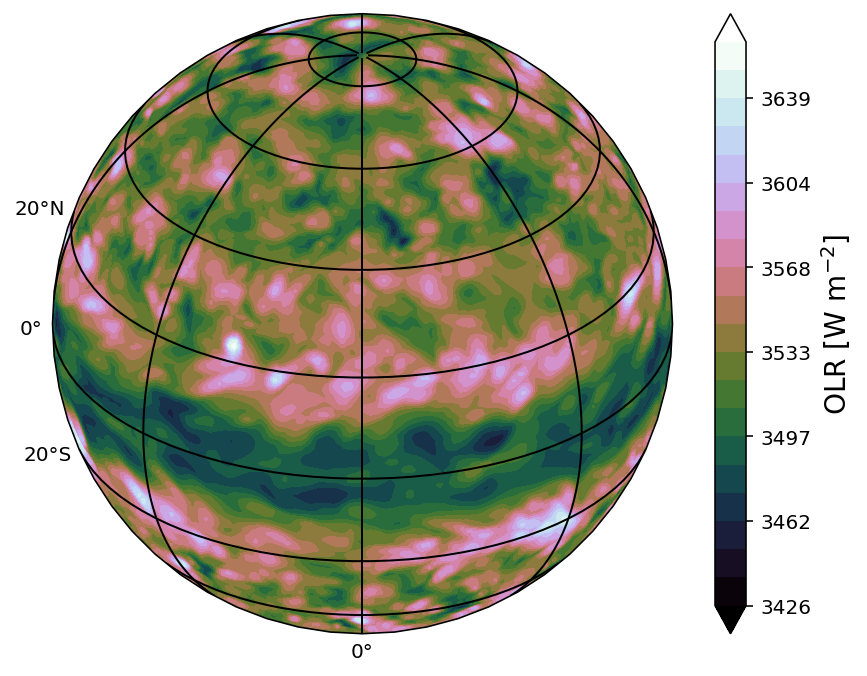}
   \includegraphics[width=0.49\textwidth]{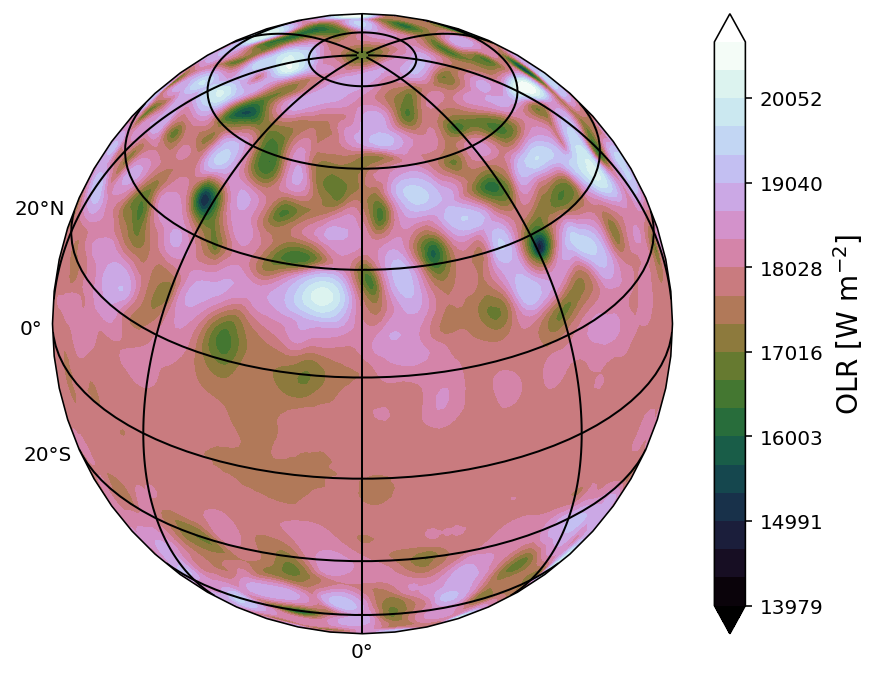}
   \includegraphics[width=0.49\textwidth]{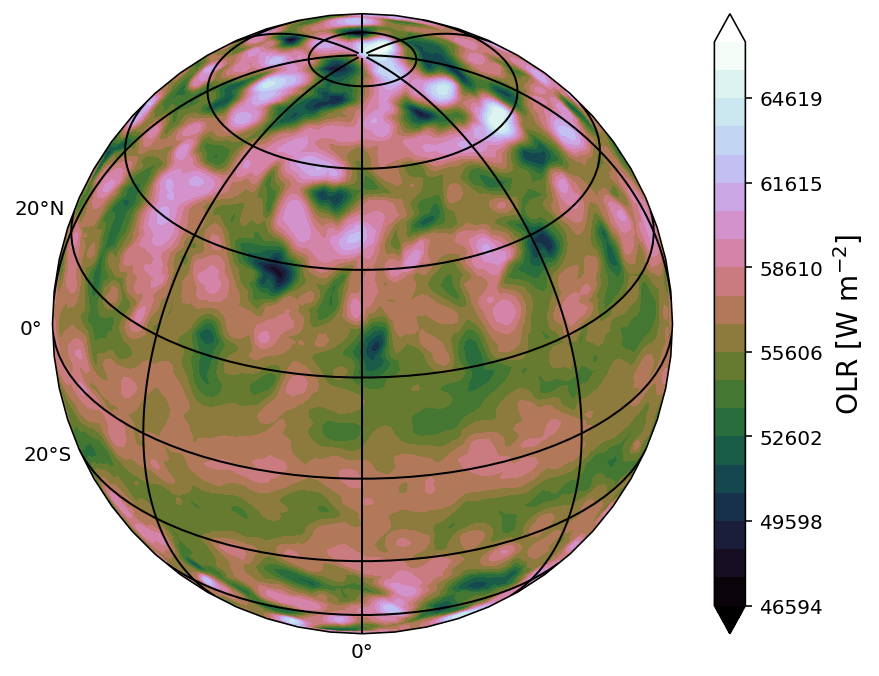}
   \includegraphics[width=0.49\textwidth]{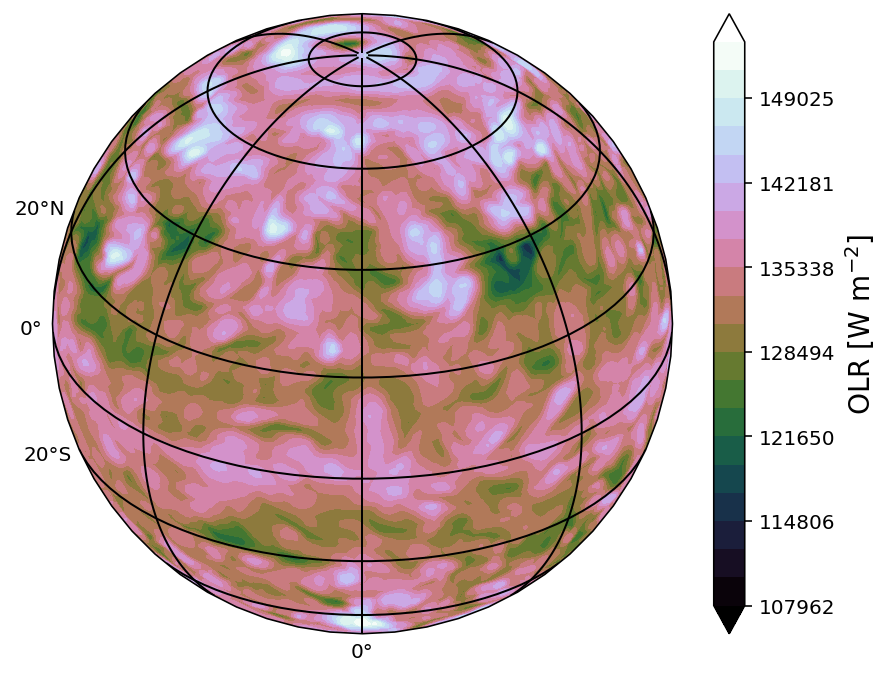}
   \includegraphics[width=0.49\textwidth]{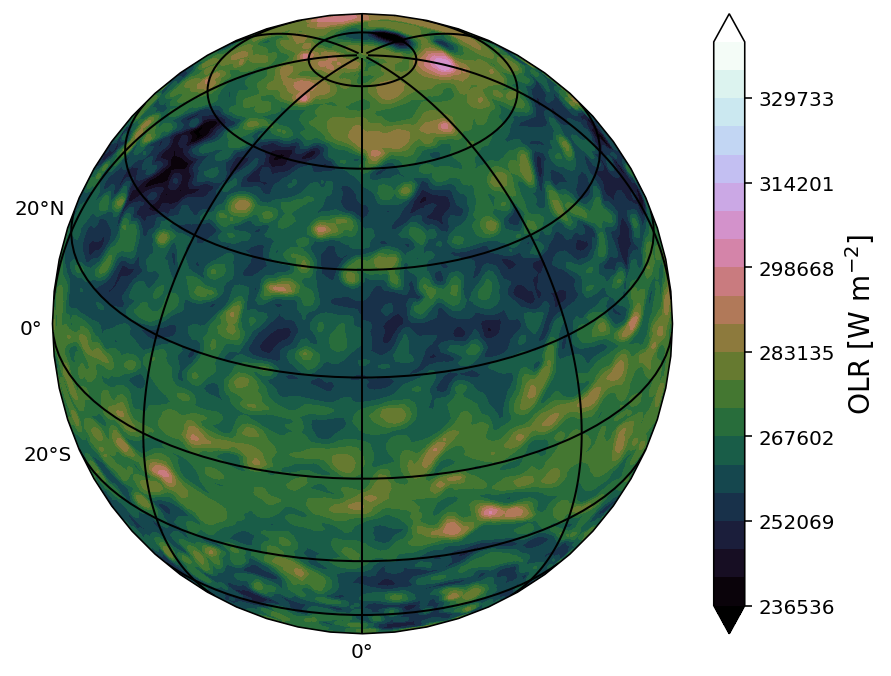}
   \caption{Outgoing longwave radiation (OLR) from each brown dwarf model at the end of the simulation. 
   T$_{\rm eff}$ = 500 K (top left), 750 K (top right), 1000 K (middle left), 1250 K (middle right) and 1500 K (bottom).
   The OLR maps show the salient dynamical features that are present in the photospheric regions of the atmosphere.
   The photosphere in these cases is at around the 1-10 bar pressure level.
   }
   \label{fig:olr}
\end{figure*}

To better elucidate the energy transport and thermal structure properties at the photosphere in the simulated brown dwarf atmospheres, we present the outgoing longwave radiation (OLR) maps in Figure \ref{fig:olr}.
Here we see that each simulation exhibits different dominant dynamical structures.
For example, the stronger thermal perturbation amplitude models show more `spottiness' and contrasts between hot and cold spots.
The T$_{\rm eff}$ = 500 K model shows the most obvious smaller-scale dynamical patterns, with clearer delineations between jet regions.
This model also shows signs of instabilities and interactions near the central jet regions.

We find in general, for the stronger amplitude forced simulations, the storm regions at higher latitudes alter the OLR patterns at the rate of the storm timescale as expected. 
The dominant dynamical features such as the equatorial and at latitude jets show less variations with time, indicating that the dynamical flows provide more thermal homogeneity and stability there.

\section{3D chemical structures}
\label{sec:3Dchem}

In this section, we examine the resulting 3D chemical structures of the coupled mini-chem GCM simulations. 
We present 1D global means and spacial maps of chemical species.

\subsection{Global means}

\begin{figure*} 
   \centering
   \includegraphics[width=0.49\textwidth]{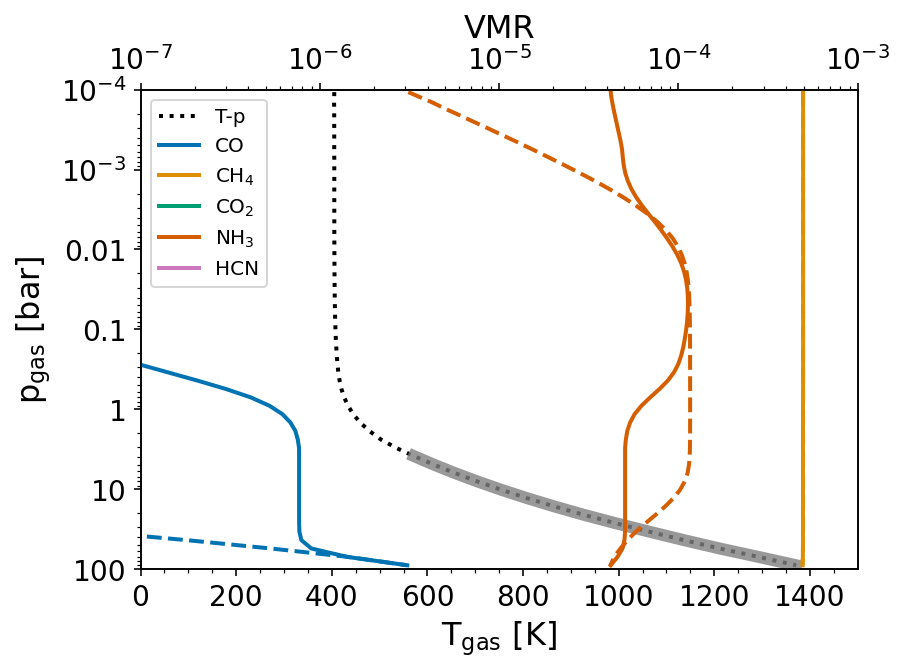}
   \includegraphics[width=0.49\textwidth]{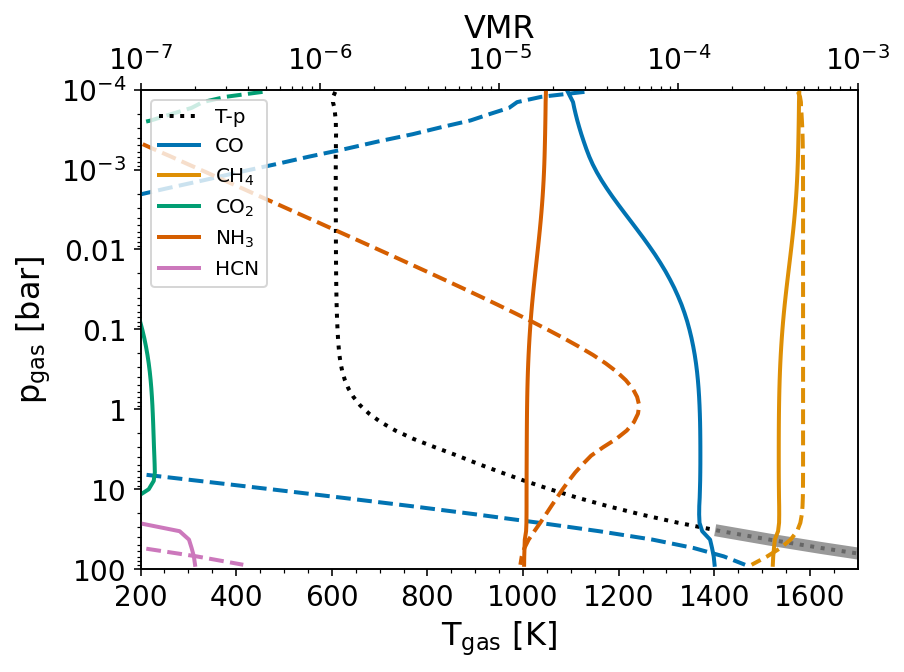}
   \includegraphics[width=0.49\textwidth]{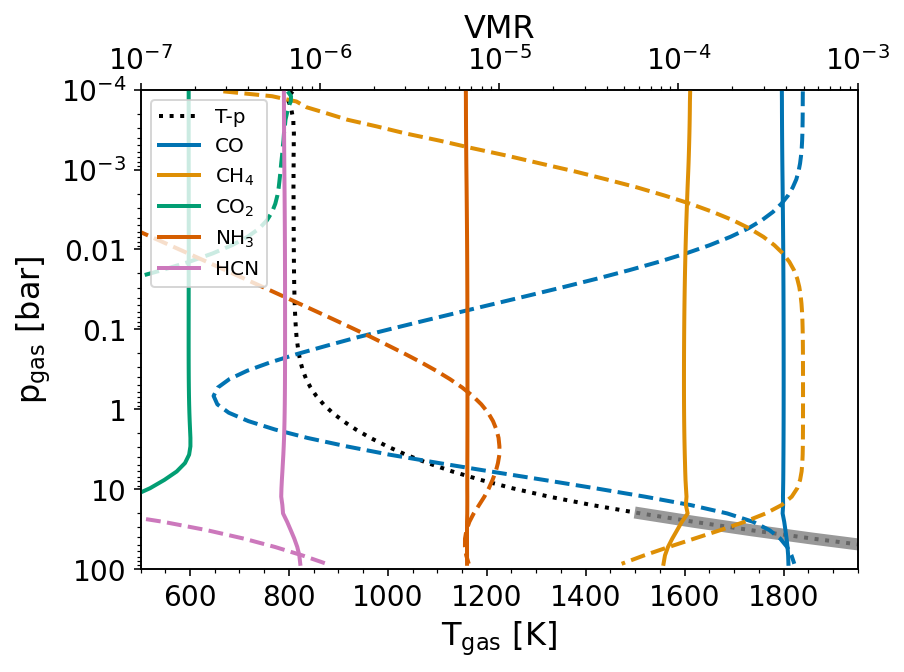}
   \includegraphics[width=0.49\textwidth]{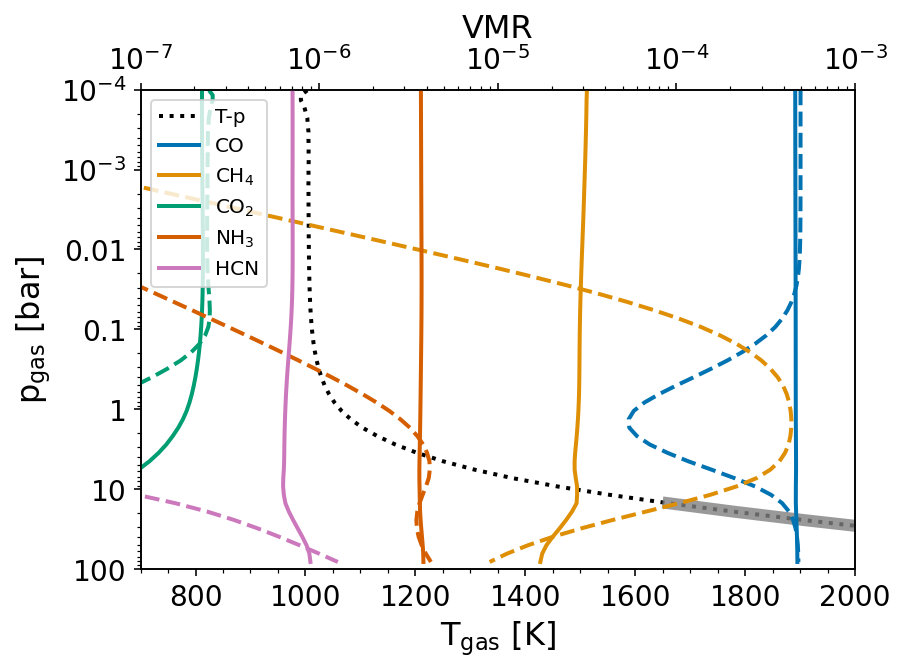}
   \includegraphics[width=0.49\textwidth]{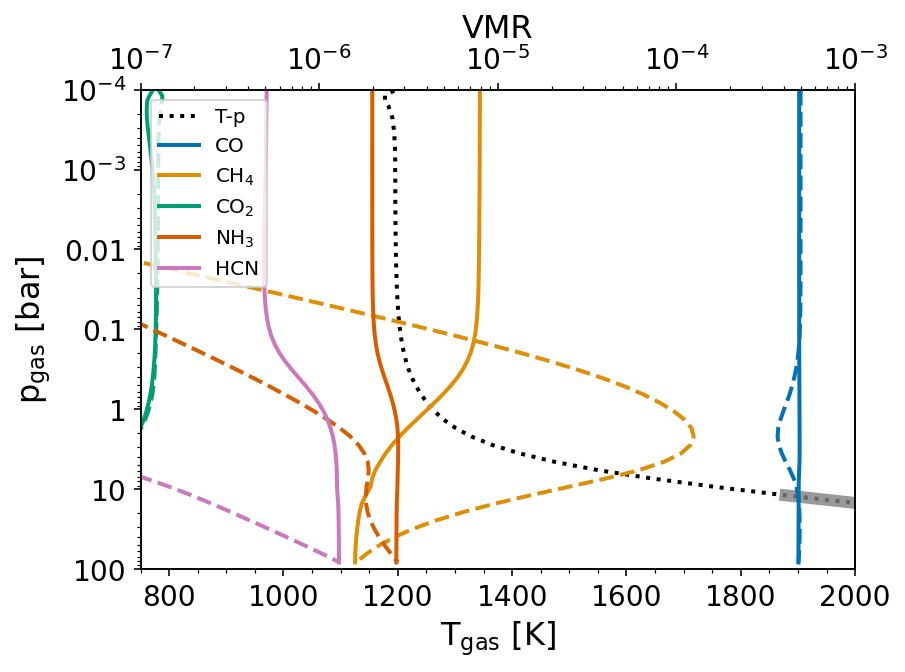}
   \caption{Global 1D mean temperature-pressure (T-p) profiles for each of the simulations (dotted black line), T$_{\rm eff}$ = 500 K (top left), 750 K (top right), 1000 K (middle left), 1250 K (middle right) and 1500 K (bottom), with the deep adiabatic region denoted by the thicker grey line. 
   Values are taken from the average results for the last 100 days of simulation.
   The global 1D mean volume mixing ratios (VMR) of various chemical species calculated with the mini-chem coupled model (coloured solid lines) and assuming chemical equilibrium (coloured dashed lines) are also overplotted.  
   }
   \label{fig:means}
\end{figure*}

\begin{figure} 
   \centering
   \includegraphics[width=0.49\textwidth]{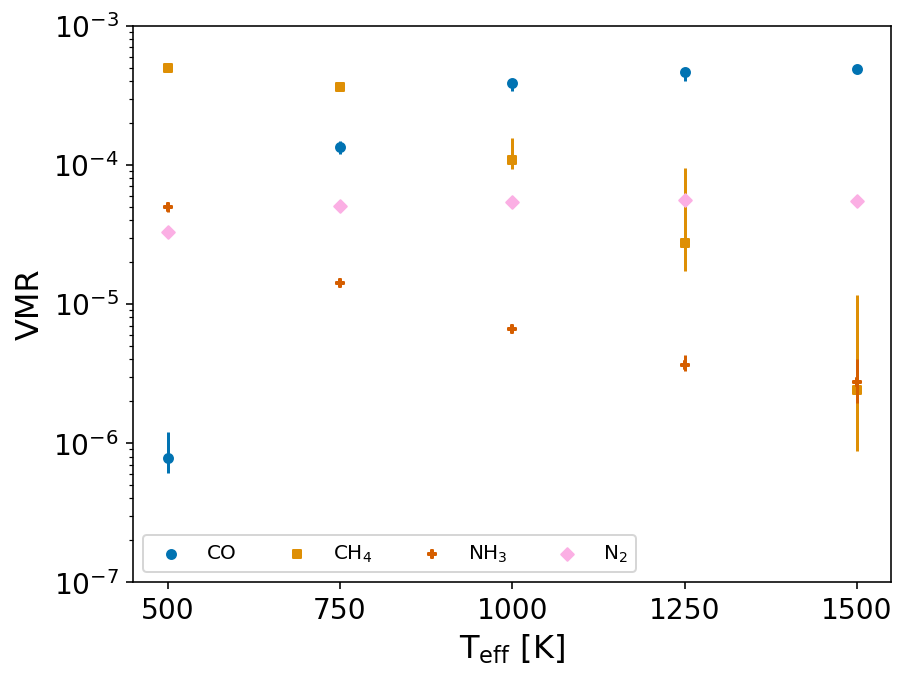}
   \caption{Mean volume mixing ratio of different species (points) at around 6 bar for each simulation, with errorbars denoting the minimum and maximum VMR range.
   Values are taken as a snapshot at the end of the simulation time.}
   \label{fig:VMR_range}
\end{figure}

In Figure \ref{fig:means} we present 1D globally averaged values of the temperature structure and volume mixing ratio (VMR) of key chemical species for each simulation, as well as the values for each species at chemical equilibrium. 
These figures also indicate the extent of the adiabatic region.
Here it is clear each brown dwarf atmosphere exhibits strong vertical mixing which leads to the non-equilibrium behaviour of the chemical species.
In particular, the strong mixing brought on by the dry convective adjustment scheme in the adiabatic regions pushes the species away from CE at the deepest parts of the atmosphere.

In Figure \ref{fig:VMR_range} we show the average VMR of CO, \ce{CH4}, \ce{NH3} and \ce{N2} along with their minimum and maximum value ranges for each effective temperature.
The trend with effective temperature of each species is seen here, with the CO-\ce{CH4} inflection point occurring at around T$_{\rm eff}$ $\approx$ 850 K, and the \ce{NH3}-\ce{N2} conversion happening at T$_{\rm eff}$ $\approx$ 550 K.
Larger variations of VMR at higher effective temperatures are attributed to the stronger thermal perturbations assumed for these models, which in turn causes stronger spatial variations in the VMR of the species.

\subsection{Comparison to VULCAN}

\begin{figure*} 
   \centering
   \includegraphics[width=0.49\textwidth]{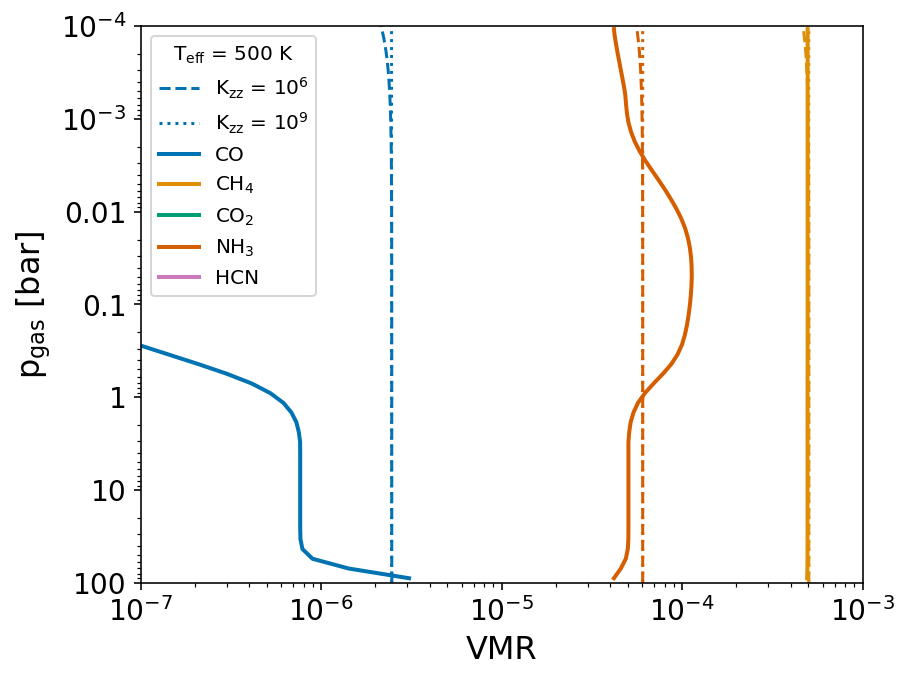}
   \includegraphics[width=0.49\textwidth]{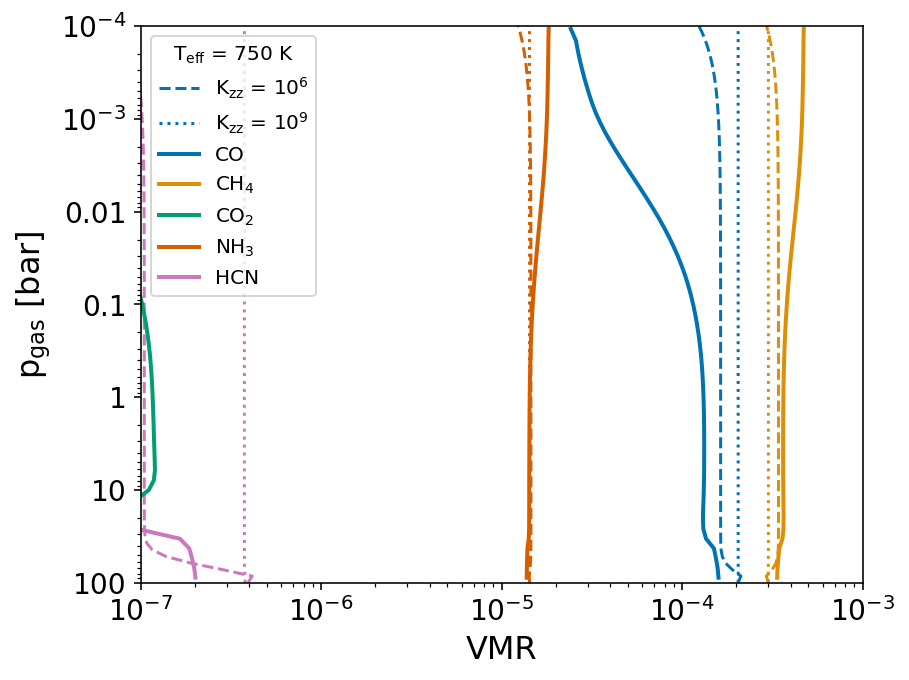}
   \includegraphics[width=0.49\textwidth]{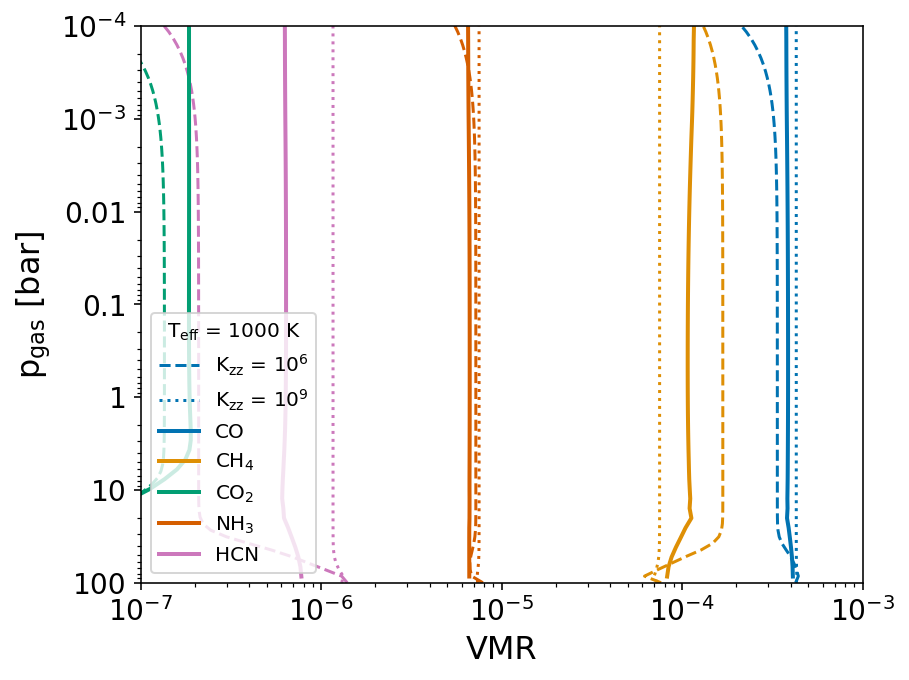}
   \includegraphics[width=0.49\textwidth]{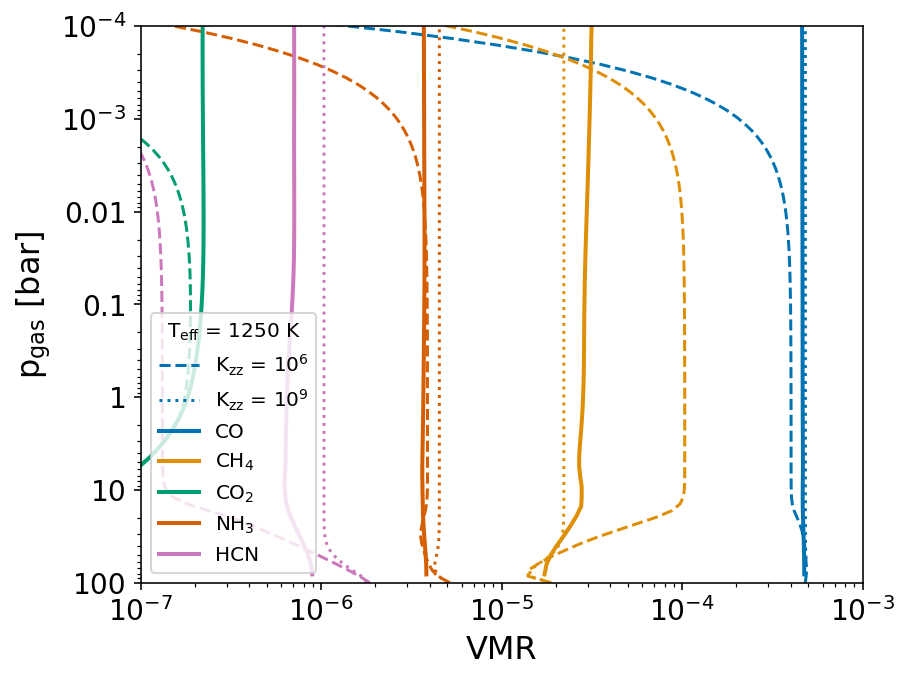}
   \includegraphics[width=0.49\textwidth]{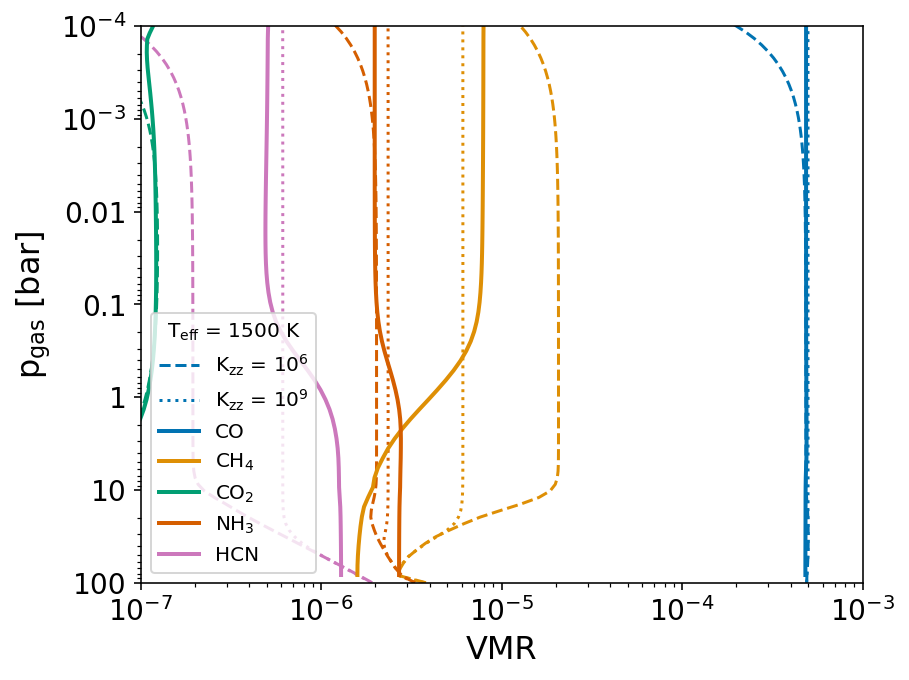}
   \caption{Global mean VMR profiles for each of the simulations (solid lines), T$_{\rm eff}$ = 500 K (top left), 750 K (top right), 1000 K (middle left), 1250 K (middle right) and 1500 K (bottom).
   Values are taken from the average results for the last 100 days of simulation.
   Dashed and dotted lines show the 1D VULCAN results using the full C-H-N-O kinetic network assuming a constant K$_{\rm zz}$ = 10$^{6}$ and  10$^{9}$ cm$^{2}$ s$^{-1}$ respectively, with the global mean T-p profile used as input.
   }
   \label{fig:v_comp}
\end{figure*}

In Figure \ref{fig:v_comp} we compare our global 1D mean VMR values from the GCM to 1D VMR profiles produced by VULCAN \citep{Tsai2017,Tsai2021} using the full C-H-N-O thermochemical network. 
We assume a constant eddy diffusion parameter of K$_{\rm zz}$ = 10$^{6}$ and  10$^{9}$ cm$^{2}$ s$^{-1}$ for the VULCAN models, taking the global mean GCM T-p profile as the input.
Differences can be seen between the 3D and 1D approaches in the quench level of some species across each effective temperature which we attribute to the K$_{\rm zz}$ approach used in VULCAN and the convective adjustment and large-scale dynamical mixing from advection inside the GCM.

This makes one-to-one comparison difficult, with many differences in the VMR structure, particularly in the deep regions, seen between the two approaches.
This suggests that possibly simulating a deeper pressure to capture species that quench $>$ 100 bar and a time-dependent tracer mixing scheme in the adiabatic regions for the GCM should be considerations for future modelling efforts.
However, overall the 1D and 3D results generally produce consistent results, to within an order of magnitude or less for most cases, suggesting that the coupled 3D scheme produces reasonable and expected outputs, comparable to 1D full kinetics models.

\subsection{Spacial variation}

\begin{figure*} 
   \centering
   \includegraphics[width=0.49\textwidth]{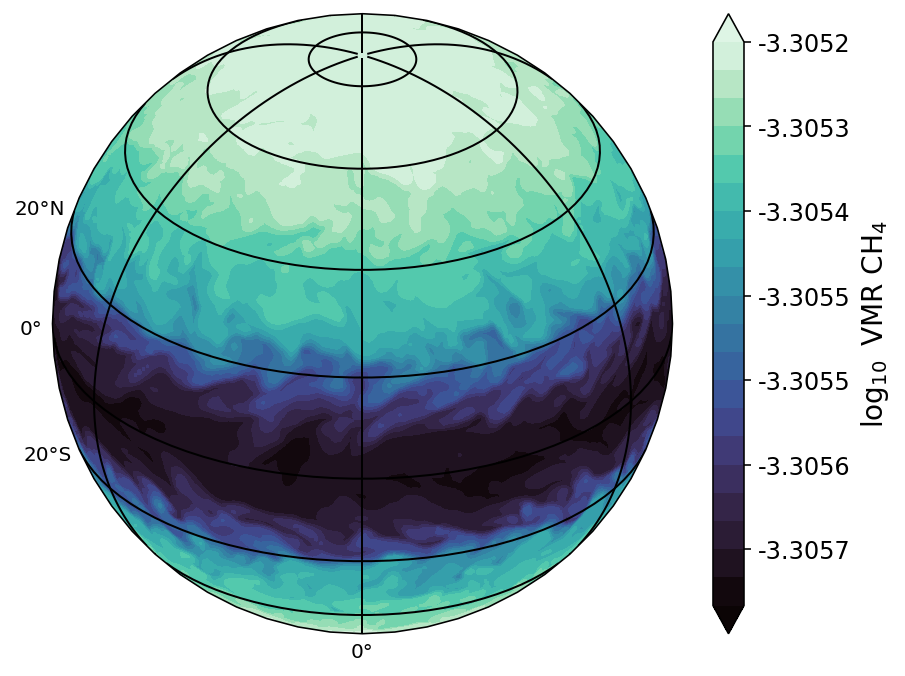}
   \includegraphics[width=0.49\textwidth]{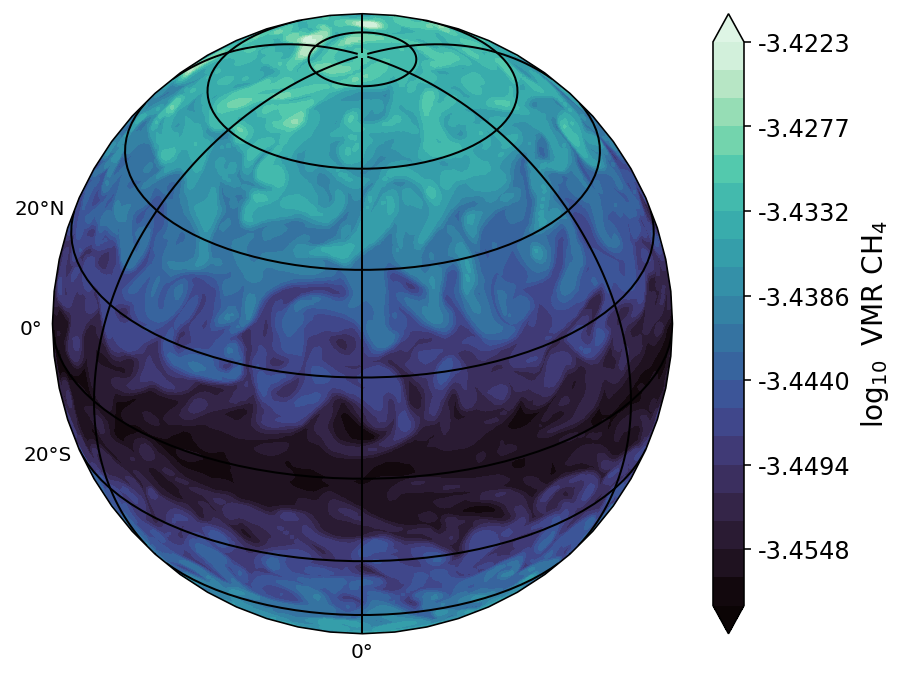}
   \includegraphics[width=0.49\textwidth]{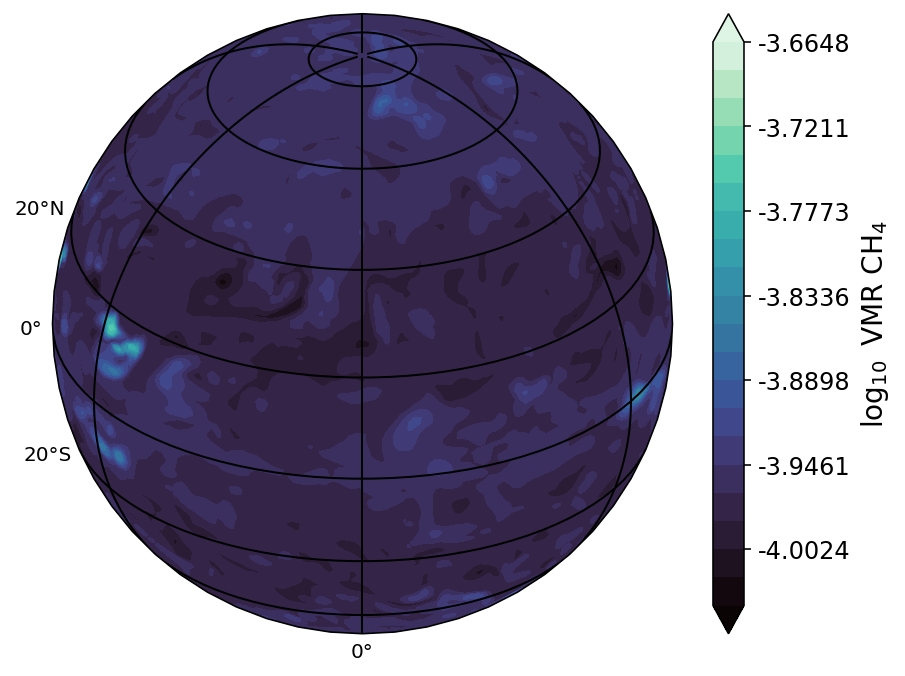}
   \includegraphics[width=0.49\textwidth]{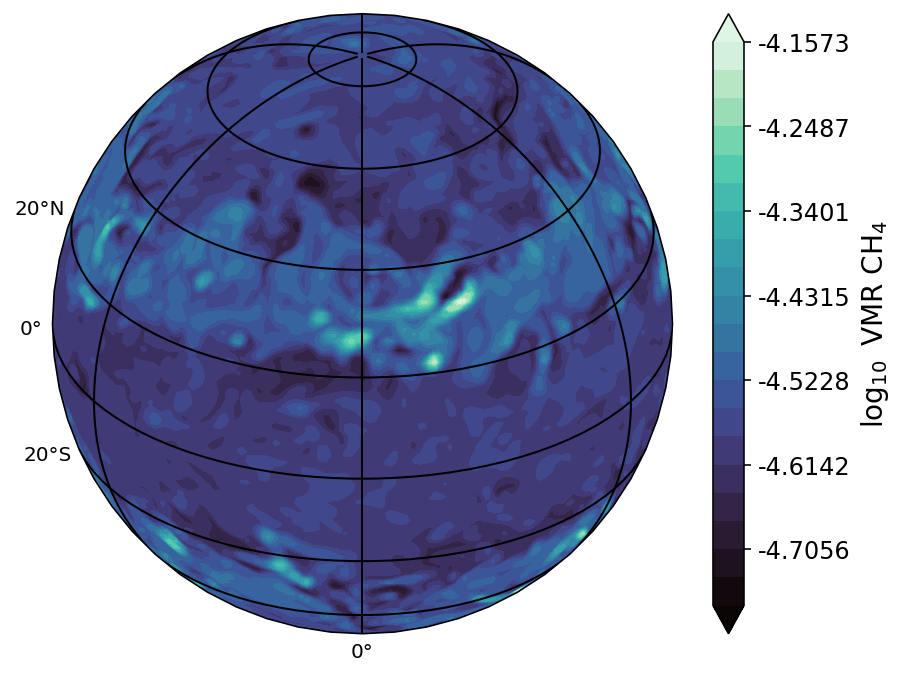}
   \includegraphics[width=0.49\textwidth]{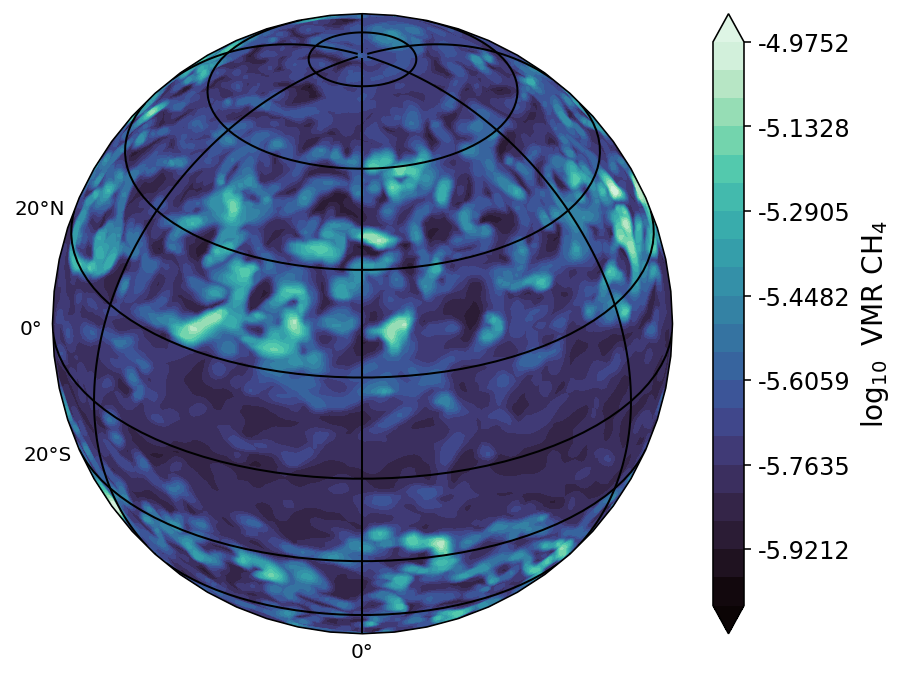}
   \caption{$\log_{10}$ volume mixing ratio (VMR) of \ce{CH4} ( colour bar) from each brown dwarf model at the end of the simulation at around a pressure level of 6 bar. 
   T$_{\rm eff}$ = 500 K (top left), 750 K (top right), 1000 K (middle left), 1250 K (middle right) and 1500 K (bottom).
   Each simulation exhibits its spatial distribution of the species dependent on the dominant dynamical feature for each parameter regime.
   Storm regions at higher latitudes alter the VMR of \ce{CH4} over the course of a storm timescale.}
   \label{fig:VMR_CH4}
\end{figure*}

\begin{figure*} 
   \centering
   \includegraphics[width=0.49\textwidth]{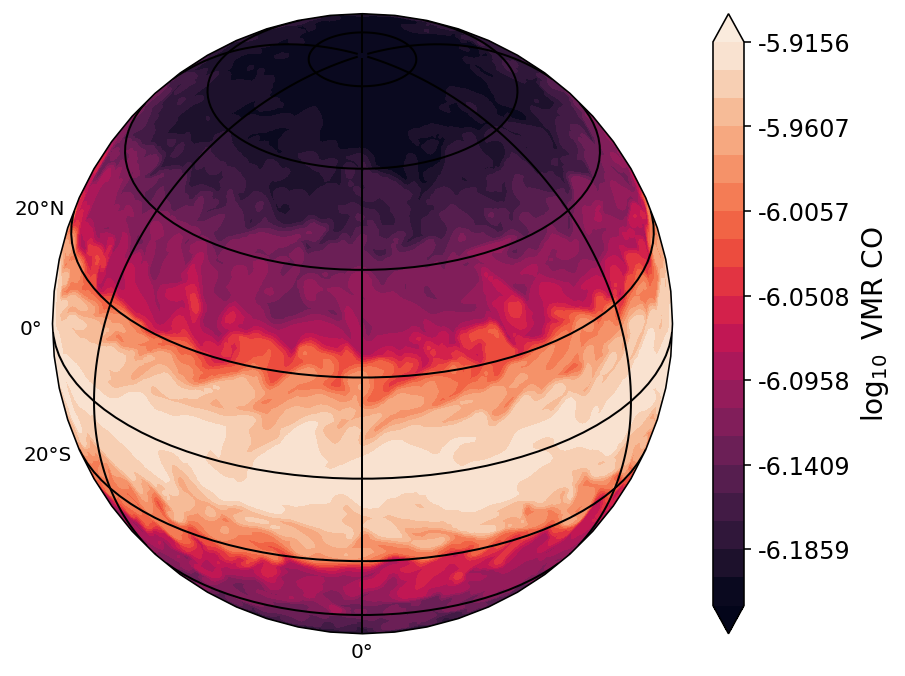}
   \includegraphics[width=0.49\textwidth]{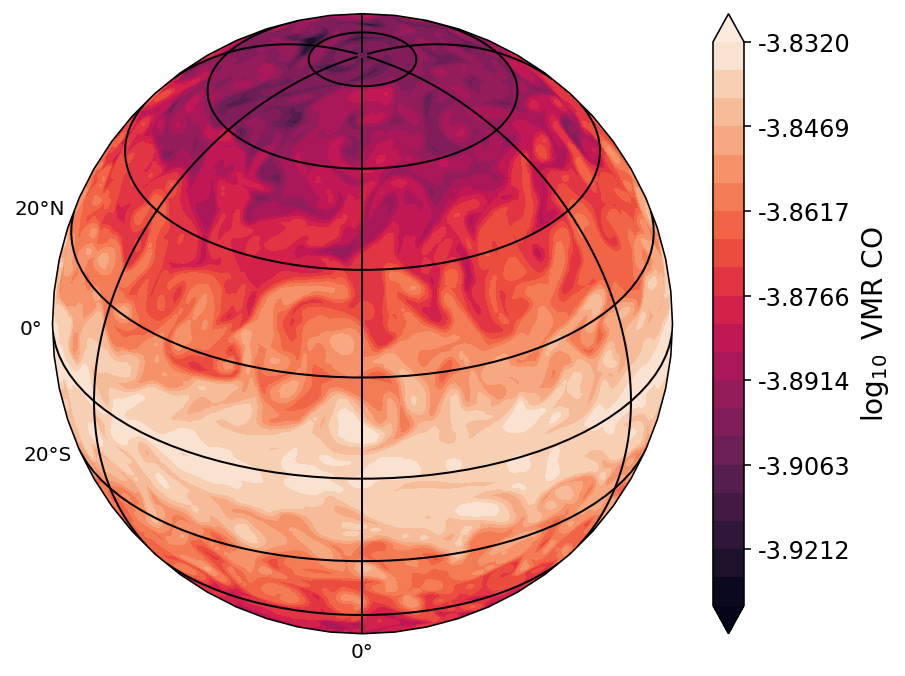}
   \includegraphics[width=0.49\textwidth]{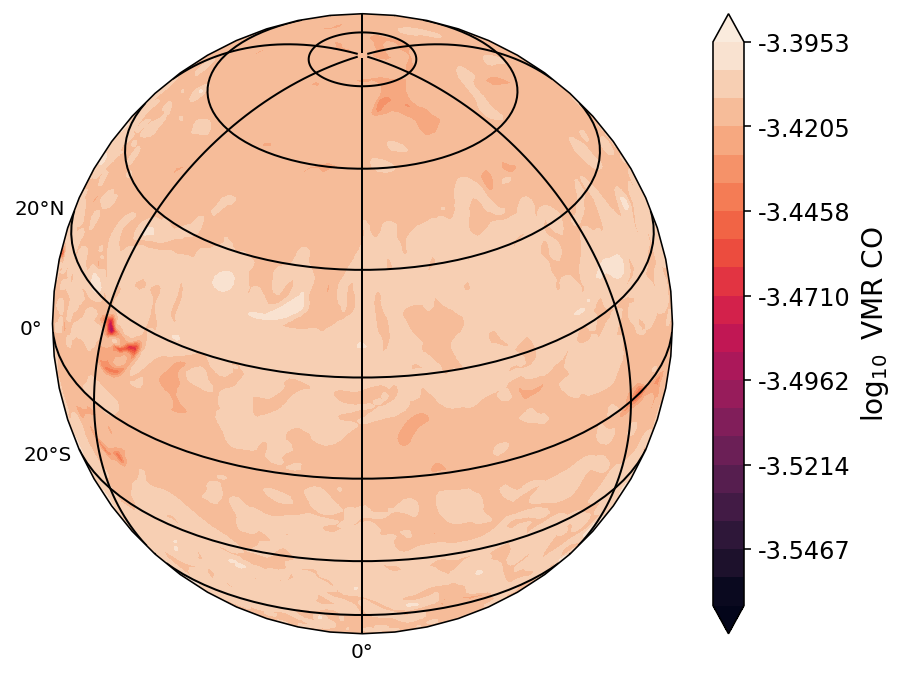}
   \includegraphics[width=0.49\textwidth]{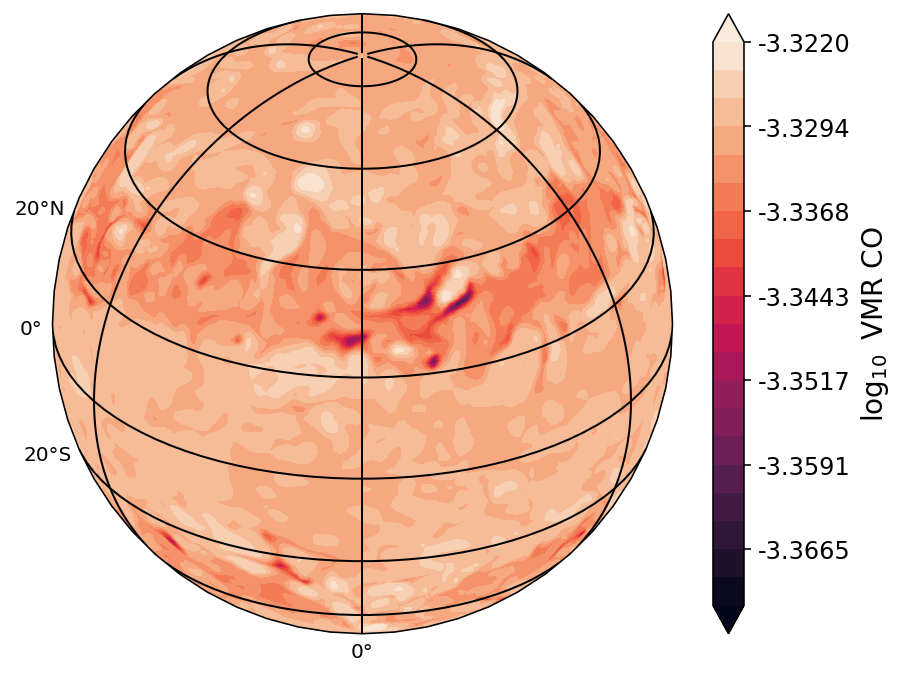}
   \includegraphics[width=0.49\textwidth]{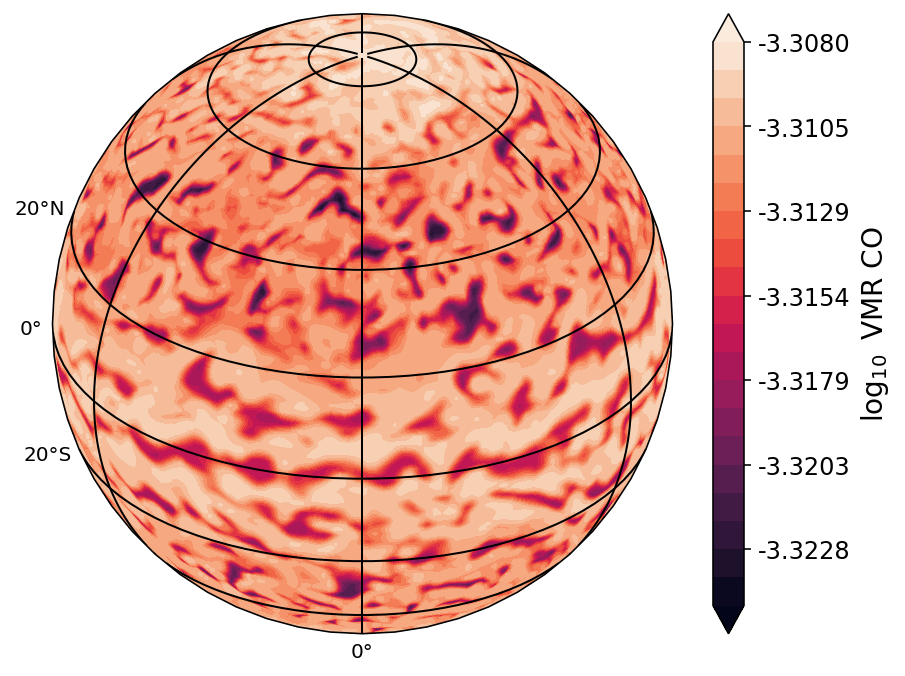}
   \caption{$\log_{10}$ volume mixing ratio (VMR) of CO (color bar) from each brown dwarf model at the end of the simulation at around a pressure level of 6 bar. 
   T$_{\rm eff}$ = 500 K (top left), 750 K (top right), 1000 K (middle left), 1250 K (middle right), and 1500 K (bottom).
   Each simulation exhibits its spatial distribution of the species dependent on the dominant dynamical feature for each parameter regime.
   Storm regions at higher latitudes alter the VMR of CO over the course of a storm timescale.}
   \label{fig:VMR_CO}
\end{figure*}

In Figure \ref{fig:VMR_CH4} and \ref{fig:VMR_CO} we show the VMR spacial distribution of \ce{CH4} and CO at a pressure level of $\approx$6 bar in each simulation.
We find that generally the spatial distribution of each species follows the dynamical structures present in the atmosphere and is also induced by localised storm formation and dissipation.
For the weaker forced and colder models (T$_{\rm eff}$ = 500 K, 750 K) the VMR spacial distribution follows closely the dynamical patterns of the atmosphere, with differences between the equatorial and at latitude jet regions.
Dynamical wave features at mid-latitudes are also present in the T$_{\rm eff}$ = 500 K and 750 K simulations, where the counter-rotating equatorial jet and pro-rotating mid-latitude jet meet.
Therefore, the main feature of these two models is that the global distribution of chemical species is highly dependent on the global scale dynamical properties of the atmosphere.
However, the absolute range of these dynamical perturbations on the VMR of the species is very small, with a highly homogeneous VMR, with less than 1\% variations, seen across the globe in these cases.

We also find that the spatial variation depends between the equator and at latitude for the hotter models (T$_{\rm eff}$ = 1000 K, 1250 K, 1500 K), again following the large-scale global dynamical patterns present in the atmosphere.
However, another key dynamical feature of these models is the formation and dissipation of storm regions, most strongly seen at higher latitudes.
We find these local storm regions provide a significant source of spatial inhomogeneity of chemical species, either reducing or increasing the influence of a species dependent on the positive or negative temperature fluctuations in the atmosphere.
These storm regions form and dissipate across a period of a given storm timescale, the `severity' and scale of the storms on the VMR of each species also depends on the amplitude of the thermal perturbation, with the stronger and hotter models showing larger VMR differences in storm regions, in line with the strength of the thermal perturbation.
These storms can induce VMR variations up to around 100\% in local regions compared to the background mean.

Overall, we find the atmospheric spacial distributions to be non-constant and variable with time and dependent on both the global dynamical structures of the brown dwarf atmosphere and the localised temperature and dynamical changes induced by the thermal perturbation scheme.
As the thermal perturbation is increased in strength, the variation in VMR within the storm regions also gets stronger.

\section{Emission spectra}
\label{sec:3Dspec}

We post-process the results of the brown dwarf models and produce emission spectra using the 3D radiative-transfer model gCMCRT \citep{Lee2022}. 
For the gas phase opacities, we include \ce{H2O} \citep{Polyansky2018}, CO \citep{Li2015}, \ce{CO2} \citep{Yurchenko2020}, \ce{CH4} \citep{Hargreaves2020}, \ce{C2H2} \citep{Chubb2020}, \ce{NH3} \citep{Coles2019} and HCN \citep{Harris2006, Barber2014}, where the line-lists used to generate the opacities are the associated reference for each molecule.
A correlated-k scheme is used at a resolution of R = 100 from 1-30 $\mu$m, with the same Gaussian sampling scheme as \citet{Marley2021}.
K-tables are combined using random overlap with resort and rebinning \citep[e.g.][]{Amundsen2017}.
\ce{H2}-\ce{H2} and \ce{H2}-He collisional induced opacity is included from \citet{Karman2019}.
The local VMR of each species is taken directly from the coupled mini-chem GCM output.

\subsection{Field of view spectra}

\begin{figure} 
   \centering
   \includegraphics[width=0.49\textwidth]{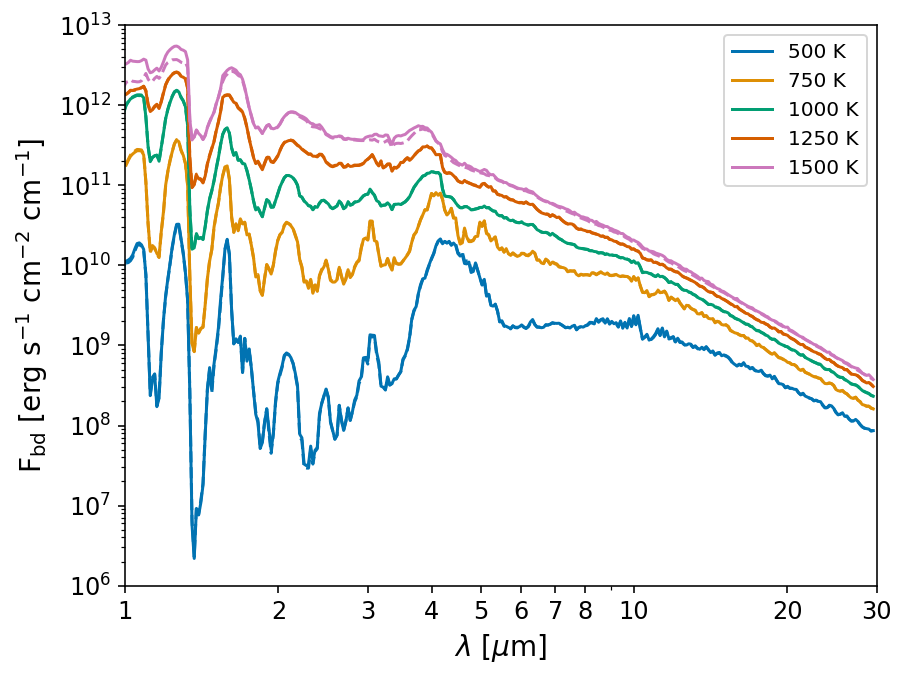}
   \caption{Emission spectra at the equator (solid lines) compared to the polar regions (dashed lines) for our different models. 
   There is no appreciable difference in the spectra between the equator and pole regions in our simulations, apart from the hottest, T$_{\rm eff}$ = 1500 K, model.}
   \label{fig:eq2pole}
\end{figure}

In Figure \ref{fig:eq2pole} we compare the spectra at the equator and pole of each brown dwarf simulation taken as a snapshot at the last output of the simulations. 
Only the T$_{\rm eff}$ = 1500 K model shows some differences in the \ce{CH4} bands in the near-infrared and 3.3 $\mu$m band.
The rest show no appreciable difference between the spectra at the equatorial or polar regions. 
This result suggests that the thermal structure and chemical composition are, on average, highly similar between the equatorial regions and the poles.
This is perhaps not too surprising as the thermal perturbations at the radiative-convective boundary are isotropic and there is no radiative feedback onto the thermal structure due to the changing chemical composition, which would also induce a perturbation to the temperature-pressure profile.
However, including such feedback is beyond the scope of the current models at this time.
We also suggest that the spatial differences in the VMR of species induced by storms at higher latitudes possibly cancel out well (at least for this one instance in time), compared to the global mean, with generally equal regions of low and high VMR of species induced by the storms.

\subsection{Rotational variability and comparison to 1D models}

\begin{figure*} 
   \centering
   \includegraphics[width=0.49\textwidth]{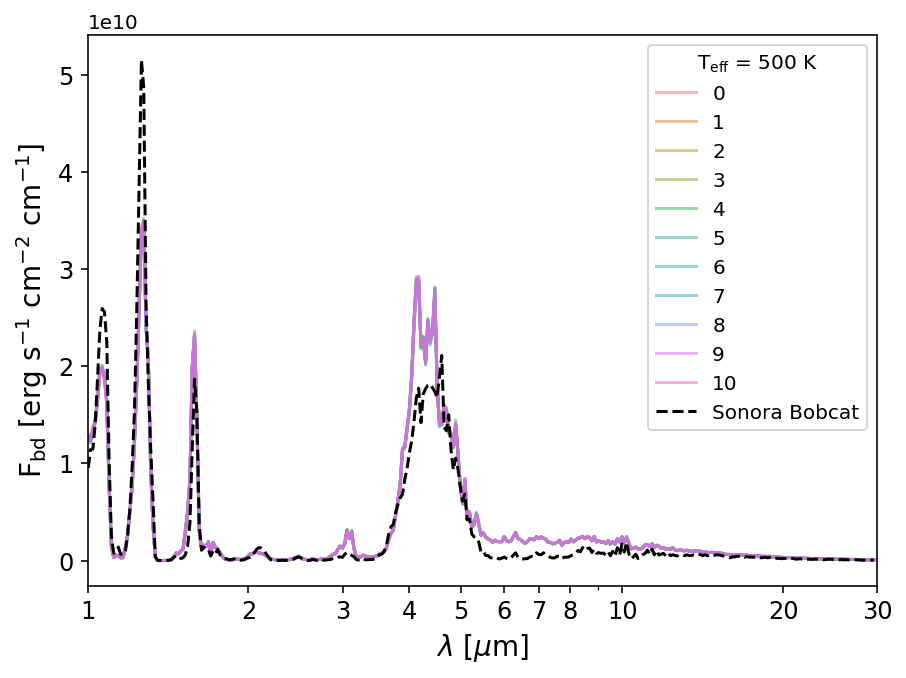}
   \includegraphics[width=0.49\textwidth]{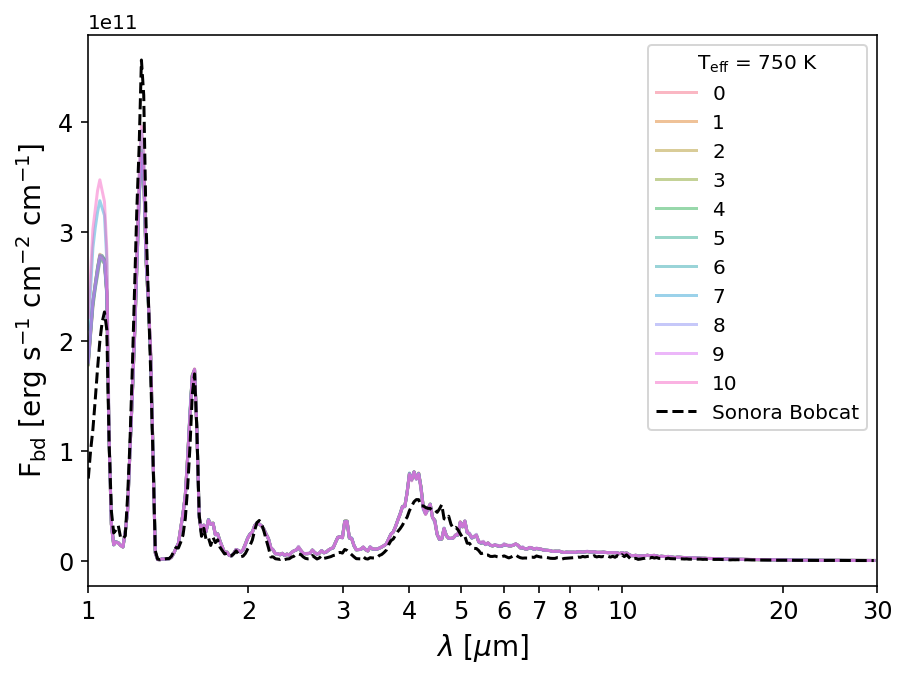}
   \includegraphics[width=0.49\textwidth]{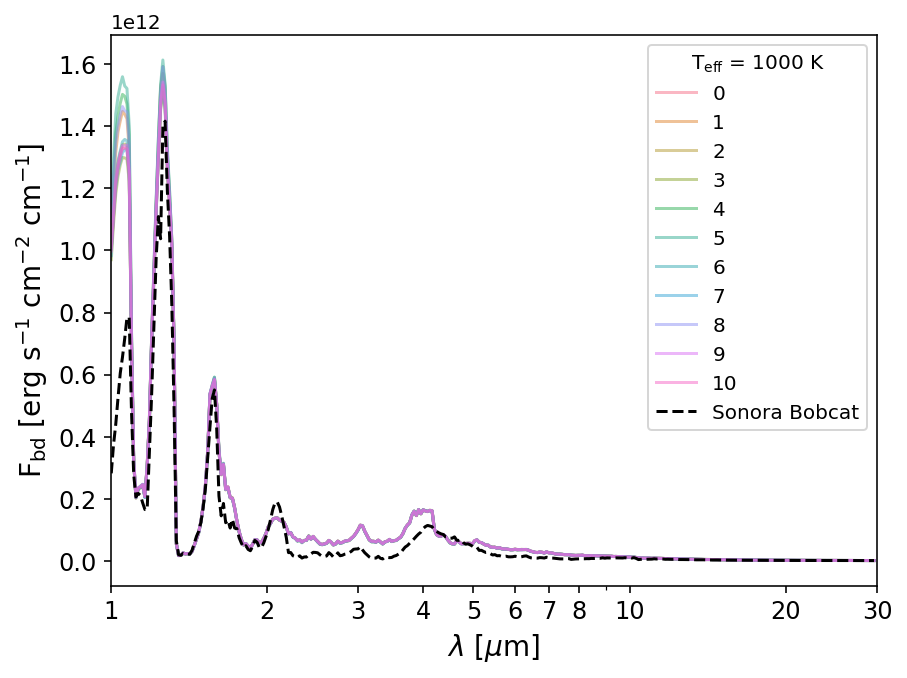}
   \includegraphics[width=0.49\textwidth]{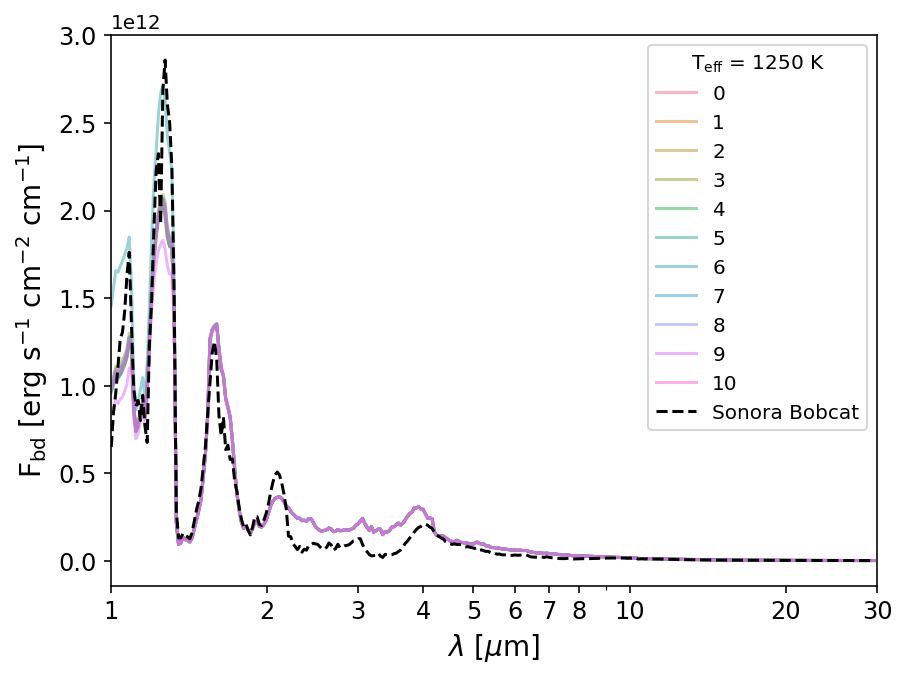}
   \includegraphics[width=0.49\textwidth]{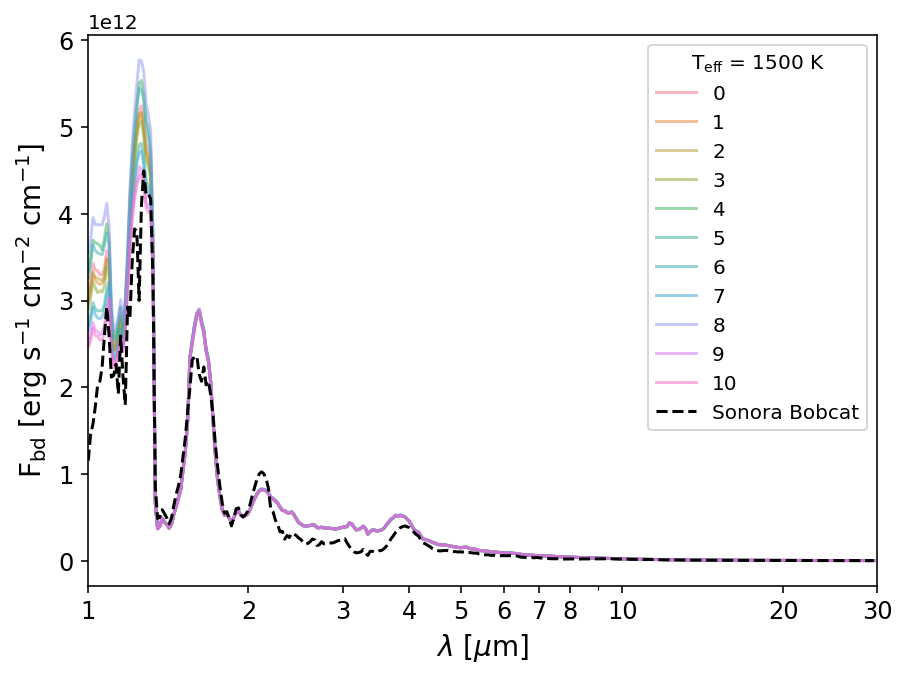}
   \caption{Emission spectra at the equator (coloured solid lines) of the brown dwarf simulations at hour intervals beyond the simulation end time for one rotation rate of 10 hours.
   T$_{\rm eff}$ = 500 K (top left), 750 K (top right), 1000 K (middle left), 1250 K (middle right) and 1500 K (bottom).
   We compare to the equivalent \citet{Marley2021} Sonora Bobcat 1D radiative-convective equilibrium models (black dashed lines).}
   \label{fig:SB}
\end{figure*}

In Figure \ref{fig:SB} we present spectra every hour of simulation past the end time for one rotation rate of 10 hours for each of our brown dwarf models.
We also take into account the rotation of the brown dwarf by calculating the emission spectrum at a longitude of +36$\degree$ each hour.
We assume an equatorial field of view of the brown dwarf, since our results in Figure \ref{fig:eq2pole} suggest overall variations in latitude are minimal.
This elucidates the inherent variability of the brown dwarf model due to fluctuations in temperature, dynamical features and VMR of chemical species from the kinetic scheme across one rotational period.

Our results show that variability over one rotational period is only present in substantial quantities at near-infrared wavelengths, in the 1-1.5 $\mu$m range, and increases with larger effective temperatures.
This is most likely due to these wavelengths probing the deeper regions that experience the greatest effects of the thermal perturbation scheme.
The lack of variation in the mid-infrared suggests that the upper atmosphere is very stable thermally and that the spatial and time variations in VMR of species do not affect the spectrum, and storms and eddies that reduce or increase the spacial VMR tend to cancel out across the hemispheres of the brown dwarfs at these lower pressure levels.

However, we note this stability comes in part due to a lack of radiative feedback onto the thermal structure from the varying VMR structures due to using our grey RT scheme.
Clouds will also affect variability significantly through impacting the thermal structure \citep{Tan2021,Tan2022}, which are not included here.
Inferences of storm formation and possible spots on brown dwarf atmospheres across the L and T range have been observed by several studies on brown dwarf variability \citep[e.g.][]{Radigan2014,Metchev2015, Apai2017, Manjavacas2019}.
Encouragingly, the dynamical properties of our simulations show similar spatial characteristics to the models proposed by \citet{Apai2017} to fit the variability seen in their brown dwarf sample.
This suggests taking into account the thermal feedback effects will be a vital consideration to model the mid-infrared wavelength variability seen in brown dwarfs.

In addition, in Figure \ref{fig:SB} we compare the equatorial spectra of each of our brown dwarf simulations to the equivalent 1D radiative-convective equilibrium Sonora Bobcat models from \citet{Marley2021} (i.e. the same T$_{\rm eff}$, gravity and metallicity as our GCM simulations).
The \citet{Marley2021} models assume chemical equilibrium with rainout when producing the 1D T-p profiles.
The output spectra from the \citet{Marley2021} are convolved to the same R = 100 wavelength grid used in gCMCRT.

Our simulated spectra compare remarkably well to the \citet{Marley2021} models considering the simplified grey radiative-transfer scheme used in the GCM model.
However, several important differences are useful to point out. 
The GCM models tend to produce flatter spectra in the regions 2-4 $\mu$m and 5-10 $\mu$m, probably due to their much more isothermal upper atmospheres compared to the correlated-k scheme used in \citet{Marley2021}. 
In addition, the GCM models generally produce a higher flux than the \citet{Marley2021} models across the infrared wavelength regime, suggesting a greater temperature present in the simulations in the upper atmosphere, probably due again to the grey opacity assumption.
Conversely, the \citet{Marley2021} models generally show a generally larger flux in the near-infrared regime, suggesting a slightly higher deeper temperature structure compared to our grey models.
The \citet{Marley2021} models also include potassium in the opacity mix, not included here, which is the primary opacity source responsible for the drop off in flux bluer than $\approx$ 1 $\mu$m.
The flux around this wavelength range from the GCM brown dwarf models is therefore likely overestimated from this lack of potassium absorption.

\section{Discussion}
\label{sec:disc}

There has been evidence suggesting that the near-IR color and variability amplitudes of brown dwarfs correlate with the viewing angles \citep[e.g.][]{vos2017,vos2022}. We have tested whether the horizontal variations of chemical species resulting from the 3D dynamics could significantly contribute to the viewing-angle-dependent spectrum, and we found that the effect is  not significant (see Figure \ref{fig:eq2pole}). This is because the zonal-mean equator-to-pole variations of chemical species in our GCMs are small. But our current results do not exclude the role of chemically driven variations as a potential mechanism because our model does not include the radiative feedback of chemical species to the dynamics. Similar to the cloud-radiative-feedback dynamics \citep{Tan2021,Tan2021b}, such a full dynamics-tracer-feedback coupling would be required to fully access potential mechanisms driving the  large equator-to-pole differences, and we leave this as a next step.

There is an interesting transition in the zonal jet streams that the cooler models exhibit multiple alternating and weak jets while the hotter models show fewer, broader, and stronger jets (see Figure \ref{fig:zonal}). 
Briefly speaking, the meridional width of the zonal jets forced by small-scale stochastic perturbations tend to follow the Rhines scale $L_{\beta}\sim \pi \sqrt{2U/\beta}$ \citep{rhines1975}, where $U$ is the magnitude of the horizontal wind speed, $\beta={\rm d}f/{\rm d}y$ is the meridional gradient of $f$ and $f$ is the Coriolis parameter.
Given a rotation rate and a planetary radius, the stronger wind speeds in our hotter models naturally lead to broader and fewer jets than in the cooler models. Figure 5 in \citet{Tan2022} also nicely illustrates this jet behaviour.
For a more in-depth discussion about the atmospheric dynamics under this forcing framework, readers are referred to our previous work of \citet{Showman2019} and \citet{Tan2022}.
Then the question comes as to why hotter models have stronger wind speeds, and this is related to how strongly we apply the thermal perturbations \citep{Tan2022}.
The choices of our forcing amplitudes for each effective temperature  are physically motivated which follows the scaling analysis in the Appendix of \citet{Showman2019}.
At a given effective temperature, if one treats the forcing amplitude  as a free parameter as in \citet{Tan2022}, the jets and the vertical mixing of chemical species would be stronger for higher forcing amplitudes.
Lastly, the vertical mixing of chemical species is effectively done by the small-scale turbulent motions rather than by the mean meridional motions associated with the jets (see the quantitative discussion in Section 4.1 of \citet{Tan2022}).
This is why in Figures \ref{fig:VMR_CH4} and \ref{fig:VMR_CO}, spatial distributions of chemical species show weak correlations to the zonal jets in Figure \ref{fig:zonal} (i.e strong chemical gradients are not cleanly delineated between the different zonal flow regions).

We have deliberately chosen not to include the formation of clouds as was done in previous similar models as they exhibit their own dynamical patterns and feedback mechanisms \citep[e.g.][]{Tan2021b, Tan2021}. 
In this study, we aim to isolate and understand the kinetic chemistry mechanisms without adding additional complex feedback from the presence of cloud coverage.
Clouds alter the local thermodynamic properties of the atmosphere due to their strong opacity, which could go on to affect the chemical compositions through changes in local temperature and wind patterns. 
Our GCM also does not capture the mixing caused by small-scale gravity waves driven by localized convection, as seen in small-scale resolving models that include clouds \citep[e.g.][]{Freytag2010,Lefevre2022}.
Our results could therefore be considered as a large-scale driven `minimum mixing' scenario in the stratospheric regions of brown dwarfs. 
Connecting the effect of clouds on the chemical kinetics will be an interesting prospect for future modelling efforts.

We assumed a simple grey opacity radiative-transfer scheme to approximate the flux through the atmosphere. 
A more accurate method would be to use a spectral correlated-k opacity method as is common in hot Jupiter modelling efforts, however, this would further increase the computational burden by an order of magnitude or more, in particular since a higher GCM spacial resolution must be used in the brown dwarf models due to their faster rotation.
Despite this, including mini-chem, clouds and a spectral radiative transfer to the base brown dwarf GCM model is a natural follow-on to the current study.
A possible avenue to explore is to apply faster correlated-k table mixing schemes such as adaptive equivalent extinction rather than random overlap \citep[e.g.][]{Amundsen2017}.
A further aspect of using a correlated-k scheme would be the inclusion of additional species that are optical wavelength absorbers such as Na and K and FeH.
These species' opacity would also have to be mixed as well, possibly assuming their abundances at chemical equilibrium in the absence of a kinetic chemistry scheme for these species. 

A wide variety of brown dwarf GCM parameters can be altered to explore dynamical structures across the expected brown dwarf parameter space.
In this study we focus on 10h rotation rates with `middling' bottom atmospheric drag, which is around the central values that were explored in \citet{Tan2021} and \citet{Tan2022}.
This was chosen in order to elucidate the basic behaviour of the coupled mini-chem GCM scheme in this atmospheric regime.
Further studies that expand the parameter space will be highly interesting, however, we note that higher rotation rate atmospheres will require a higher GCM spacial resolution which would hinder the efficiency of the kinetic coupled scheme significantly.

We note that some species will quench at a higher pressure than the simulation boundary used in this study, this was seen in the \citet{Zahnle2014} 1D models, where dependent on the parameter regime quenching could occur at pressures $>$ 100 bar, beyond the lower boundary of the GCM models.
We note however, that including deeper regions in the GCM will significantly increase the time for dynamical convergence of the GCM.
Improvements to the kinetic chemical scheme to be more computationally efficient at higher pressures and temperatures will also be required to complete these simulations.
This would be something to implement in future similar modelling efforts of this type in order to provide a more accurate quench level for species.

\section{Summary and conclusions}
\label{sec:conc}

In this study, we explored a further small step in a holistic understanding of the dynamical features of brown dwarf atmospheres. 
We performed idealised 3D GCM models of a range of brown dwarf effective temperatures coupled to the miniature chemical kinetic scheme mini-chem in order to examine their 3D chemical structures.
Our simulations show the rich spacial variations brought on by the interactions between the large-scale dynamical flows and small-scale atmospheric chemistry.
We reproduce the strong non-equilibrium chemical signatures expected in brown dwarf atmospheres from vertical mixing.
Generally, species are well homogenised across the globe due to vertical and horizontal mixing of the atmosphere, with mixing ratio variations less than 10\%, dependent on the species.
This is encouraging for 1D chemical kinetic models of brown dwarfs which concern the time- and global-mean state of the atmospheres. \citet{Tan2022} provided analytic calculations of the mean vertical mixing coefficient, $K_{\rm zz}$, under this forcing framework for chemical and cloud tracers and could be helpful for 1D models. 
This is further confirmed with the good agreement between the 1D VULCAN and global averaged GCM results presented in this study, which suggests 1D models may take advantage of GCM derived T-p and mixing profiles when characterising these atmospheres.

Meanwhile, the development of storms and eddies in the atmosphere can produce localised regions of depletion or replenishment of chemical species at higher contrast to the background mean, sometimes up to 100\%.
These storms form and dissipate dependent on the given storm timescale and perturbation amplitudes in the brown dwarf simulations.
In addition, similar to hot Jupiter studies, we find that the chemical structure of the atmosphere is inherently linked to the dynamical features of the brown dwarf atmosphere, with spacial variations of species following the wave patterns present in each simulation.
This 3D effect is an important consideration when interpreting the time-resolved spectroscopic variability of brown dwarfs.

Spectra wise, little variation between equatorial and polar regions is seen as well as variability over the course of a rotational rate.
Most variability is seen in the near-infrared wavelengths (1-1.5 $\mu$m), due to these wavelengths probing the thermal perturbation regions.
This lack of variability is possibly due to our grey opacity assumption and cloud-free models.
Including a cloud and correlated-k scheme able to take into account the changing spacial VMR of species produced by kinetics is a natural next step for these models.

Overall, our study elucidates further the complicated 3D coupling between chemistry and atmospheric dynamics for brown dwarf atmospheres.
This study represents a useful stepping stone to the eventual performing self-consistent 3D chemical modelling of these atmospheres in the future.

\section*{Acknowledgements}
E.K.H. Lee is supported by the SNSF Ambizione Fellowship grant (\#193448).
X. Tan was supported by the European community through the ERC advanced grant EXOCONDENSE (PI: R.T. Pierrehumbert).
S-M. Tsai is supported by the University of California at Riverside.
The HPC support staff at AOPP, University of Oxford and University of Bern is highly acknowledged.

\section*{Data Availability}
The 1D radiative-transfer, mini-chem and gCMCRT source codes are available on the lead author's GitHub: \url{https://github.com/ELeeAstro}.
GCM output in NetCDF format and animated GIFs of Figures \ref{fig:olr}, \ref{fig:VMR_CH4} and \ref{fig:VMR_CO} are available on Zenodo: \url{https://zenodo.org/record/7806376}.
All other data is available upon request to the lead author.


\bibliographystyle{mnras}
\bibliography{bib} 

\begin{thebibliography}{}
\makeatletter
\relax
\def\mn@urlcharsother{\let\do\@makeother \do\$\do\&\do\#\do\^\do\_\do\%\do\~}
\def\mn@doi{\begingroup\mn@urlcharsother \@ifnextchar [ {\mn@doi@}
  {\mn@doi@[]}}
\def\mn@doi@[#1]#2{\def\@tempa{#1}\ifx\@tempa\@empty \href
  {http://dx.doi.org/#2} {doi:#2}\else \href {http://dx.doi.org/#2} {#1}\fi
  \endgroup}
\def\mn@eprint#1#2{\mn@eprint@#1:#2::\@nil}
\def\mn@eprint@arXiv#1{\href {http://arxiv.org/abs/#1} {{\tt arXiv:#1}}}
\def\mn@eprint@dblp#1{\href {http://dblp.uni-trier.de/rec/bibtex/#1.xml}
  {dblp:#1}}
\def\mn@eprint@#1:#2:#3:#4\@nil{\def\@tempa {#1}\def\@tempb {#2}\def\@tempc
  {#3}\ifx \@tempc \@empty \let \@tempc \@tempb \let \@tempb \@tempa \fi \ifx
  \@tempb \@empty \def\@tempb {arXiv}\fi \@ifundefined
  {mn@eprint@\@tempb}{\@tempb:\@tempc}{\expandafter \expandafter \csname
  mn@eprint@\@tempb\endcsname \expandafter{\@tempc}}}

\bibitem[\protect\citeauthoryear{{Ackerman} \& {Marley}}{{Ackerman} \&
  {Marley}}{2001}]{Ackerman2001}
{Ackerman} A.~S.,  {Marley} M.~S.,  2001, \mn@doi [\apj] {10.1086/321540},
  \href {https://ui.adsabs.harvard.edu/abs/2001ApJ...556..872A} {556, 872}

\bibitem[\protect\citeauthoryear{{Alderson} et~al.,}{{Alderson}
  et~al.}{2023}]{Alderson2023}
{Alderson} L.,  et~al., 2023, \mn@doi [\nat] {10.1038/s41586-022-05591-3},
  \href {https://ui.adsabs.harvard.edu/abs/2023Natur.614..664A} {614, 664}

\bibitem[\protect\citeauthoryear{{Allard}, {Hauschildt}, {Alexander}, {Tamanai}
   \& {Schweitzer}}{{Allard} et~al.}{2001}]{Allard2001}
{Allard} F.,  {Hauschildt} P.~H.,  {Alexander} D.~R.,  {Tamanai} A.,
  {Schweitzer} A.,  2001, \mn@doi [\apj] {10.1086/321547}, \href
  {https://ui.adsabs.harvard.edu/abs/2001ApJ...556..357A} {556, 357}

\bibitem[\protect\citeauthoryear{{Allard}, {Homeier}  \& {Freytag}}{{Allard}
  et~al.}{2011}]{Allard2011}
{Allard} F.,  {Homeier} D.,   {Freytag} B.,  2011, in {Johns-Krull} C.,
  {Browning} M.~K.,   {West} A.~A.,  eds,  Astronomical Society of the Pacific
  Conference Series Vol. 448, 16th Cambridge Workshop on Cool Stars, Stellar
  Systems, and the Sun. p.~91 (\mn@eprint {arXiv} {1011.5405}),
  \mn@doi{10.48550/arXiv.1011.5405}

\bibitem[\protect\citeauthoryear{{Amundsen}, {Tremblin}, {Manners}, {Baraffe}
  \& {Mayne}}{{Amundsen} et~al.}{2017}]{Amundsen2017}
{Amundsen} D.~S.,  {Tremblin} P.,  {Manners} J.,  {Baraffe} I.,   {Mayne}
  N.~J.,  2017, \mn@doi [\aap] {10.1051/0004-6361/201629322}, \href
  {https://ui.adsabs.harvard.edu/abs/2017A&A...598A..97A} {598, A97}

\bibitem[\protect\citeauthoryear{{Apai} et~al.,}{{Apai}
  et~al.}{2017}]{Apai2017}
{Apai} D.,  et~al., 2017, \mn@doi [Science] {10.1126/science.aam9848}, \href
  {https://ui.adsabs.harvard.edu/abs/2017Sci...357..683A} {357, 683}

\bibitem[\protect\citeauthoryear{{Barber}, {Strange}, {Hill}, {Polyansky},
  {Mellau}, {Yurchenko}  \& {Tennyson}}{{Barber} et~al.}{2014}]{Barber2014}
{Barber} R.~J.,  {Strange} J.~K.,  {Hill} C.,  {Polyansky} O.~L.,  {Mellau}
  G.~C.,  {Yurchenko} S.~N.,   {Tennyson} J.,  2014, \mn@doi [\mnras]
  {10.1093/mnras/stt2011}, \href
  {https://ui.adsabs.harvard.edu/abs/2014MNRAS.437.1828B} {437, 1828}

\bibitem[\protect\citeauthoryear{{Beltz}, {Rauscher}, {Roman}  \&
  {Guilliat}}{{Beltz} et~al.}{2022}]{Beltz2022}
{Beltz} H.,  {Rauscher} E.,  {Roman} M.~T.,   {Guilliat} A.,  2022, \mn@doi
  [\aj] {10.3847/1538-3881/ac3746}, \href
  {https://ui.adsabs.harvard.edu/abs/2022AJ....163...35B} {163, 35}

\bibitem[\protect\citeauthoryear{{Bordwell}, {Brown}  \& {Oishi}}{{Bordwell}
  et~al.}{2018}]{Bordwell2018}
{Bordwell} B.,  {Brown} B.~P.,   {Oishi} J.~S.,  2018, \mn@doi [\apj]
  {10.3847/1538-4357/aaa551}, \href
  {https://ui.adsabs.harvard.edu/abs/2018ApJ...854....8B} {854, 8}

\bibitem[\protect\citeauthoryear{{Burgasser} et~al.,}{{Burgasser}
  et~al.}{2002a}]{Burgasser2002}
{Burgasser} A.~J.,  et~al., 2002a, \mn@doi [\apj] {10.1086/324033}, \href
  {https://ui.adsabs.harvard.edu/abs/2002ApJ...564..421B} {564, 421}

\bibitem[\protect\citeauthoryear{{Burgasser}, {Marley}, {Ackerman}, {Saumon},
  {Lodders}, {Dahn}, {Harris}  \& {Kirkpatrick}}{{Burgasser}
  et~al.}{2002b}]{Burgasser2022b}
{Burgasser} A.~J.,  {Marley} M.~S.,  {Ackerman} A.~S.,  {Saumon} D.,  {Lodders}
  K.,  {Dahn} C.~C.,  {Harris} H.~C.,   {Kirkpatrick} J.~D.,  2002b, \mn@doi
  [\apjl] {10.1086/341343}, \href
  {https://ui.adsabs.harvard.edu/abs/2002ApJ...571L.151B} {571, L151}

\bibitem[\protect\citeauthoryear{{Calamari} et~al.,}{{Calamari}
  et~al.}{2022}]{Calamari2022}
{Calamari} E.,  et~al., 2022, \mn@doi [\apj] {10.3847/1538-4357/ac9cc9}, \href
  {https://ui.adsabs.harvard.edu/abs/2022ApJ...940..164C} {940, 164}

\bibitem[\protect\citeauthoryear{{Carone} et~al.,}{{Carone}
  et~al.}{2020}]{Carone2020}
{Carone} L.,  et~al., 2020, \mn@doi [\mnras] {10.1093/mnras/staa1733}, \href
  {https://ui.adsabs.harvard.edu/abs/2020MNRAS.496.3582C} {496, 3582}

\bibitem[\protect\citeauthoryear{{Chubb}, {Tennyson}  \& {Yurchenko}}{{Chubb}
  et~al.}{2020}]{Chubb2020}
{Chubb} K.~L.,  {Tennyson} J.,   {Yurchenko} S.~N.,  2020, \mn@doi [\mnras]
  {10.1093/mnras/staa229}, \href
  {https://ui.adsabs.harvard.edu/abs/2020MNRAS.493.1531C} {493, 1531}

\bibitem[\protect\citeauthoryear{{Coles}, {Yurchenko}  \& {Tennyson}}{{Coles}
  et~al.}{2019}]{Coles2019}
{Coles} P.~A.,  {Yurchenko} S.~N.,   {Tennyson} J.,  2019, \mn@doi [\mnras]
  {10.1093/mnras/stz2778}, \href
  {https://ui.adsabs.harvard.edu/abs/2019MNRAS.490.4638C} {490, 4638}

\bibitem[\protect\citeauthoryear{{Drummond} et~al.,}{{Drummond}
  et~al.}{2020}]{Drummond2020}
{Drummond} B.,  et~al., 2020, \mn@doi [\aap] {10.1051/0004-6361/201937153},
  \href {https://ui.adsabs.harvard.edu/abs/2020A&A...636A..68D} {636, A68}

\bibitem[\protect\citeauthoryear{{Fegley} \& {Lodders}}{{Fegley} \&
  {Lodders}}{1996}]{Fegley1996}
{Fegley} Bruce J.,  {Lodders} K.,  1996, \mn@doi [\apjl] {10.1086/310356},
  \href {https://ui.adsabs.harvard.edu/abs/1996ApJ...472L..37F} {472, L37}

\bibitem[\protect\citeauthoryear{{Freedman}, {Lustig-Yaeger}, {Fortney},
  {Lupu}, {Marley}  \& {Lodders}}{{Freedman} et~al.}{2014}]{Freedman2014}
{Freedman} R.~S.,  {Lustig-Yaeger} J.,  {Fortney} J.~J.,  {Lupu} R.~E.,
  {Marley} M.~S.,   {Lodders} K.,  2014, \mn@doi [\apjs]
  {10.1088/0067-0049/214/2/25}, \href
  {https://ui.adsabs.harvard.edu/abs/2014ApJS..214...25F} {214, 25}

\bibitem[\protect\citeauthoryear{{Freytag}, {Allard}, {Ludwig}, {Homeier}  \&
  {Steffen}}{{Freytag} et~al.}{2010}]{Freytag2010}
{Freytag} B.,  {Allard} F.,  {Ludwig} H.~G.,  {Homeier} D.,   {Steffen} M.,
  2010, \mn@doi [\aap] {10.1051/0004-6361/200913354}, \href
  {https://ui.adsabs.harvard.edu/abs/2010A&A...513A..19F} {513, A19}

\bibitem[\protect\citeauthoryear{{Geballe}, {Saumon}, {Golimowski}, {Leggett},
  {Marley}  \& {Noll}}{{Geballe} et~al.}{2009}]{Geballe2009}
{Geballe} T.~R.,  {Saumon} D.,  {Golimowski} D.~A.,  {Leggett} S.~K.,  {Marley}
  M.~S.,   {Noll} K.~S.,  2009, \mn@doi [\apj] {10.1088/0004-637X/695/2/844},
  \href {https://ui.adsabs.harvard.edu/abs/2009ApJ...695..844G} {695, 844}

\bibitem[\protect\citeauthoryear{{Hammond} \& {Pierrehumbert}}{{Hammond} \&
  {Pierrehumbert}}{2017}]{Hammond2017}
{Hammond} M.,  {Pierrehumbert} R.~T.,  2017, \mn@doi [\apj]
  {10.3847/1538-4357/aa9328}, \href
  {https://ui.adsabs.harvard.edu/abs/2017ApJ...849..152H} {849, 152}

\bibitem[\protect\citeauthoryear{{Hargreaves}, {Gordon}, {Rey}, {Nikitin},
  {Tyuterev}, {Kochanov}  \& {Rothman}}{{Hargreaves}
  et~al.}{2020}]{Hargreaves2020}
{Hargreaves} R.~J.,  {Gordon} I.~E.,  {Rey} M.,  {Nikitin} A.~V.,  {Tyuterev}
  V.~G.,  {Kochanov} R.~V.,   {Rothman} L.~S.,  2020, \mn@doi [\apjs]
  {10.3847/1538-4365/ab7a1a}, \href
  {https://ui.adsabs.harvard.edu/abs/2020ApJS..247...55H} {247, 55}

\bibitem[\protect\citeauthoryear{{Harris}, {Tennyson}, {Kaminsky}, {Pavlenko}
  \& {Jones}}{{Harris} et~al.}{2006}]{Harris2006}
{Harris} G.~J.,  {Tennyson} J.,  {Kaminsky} B.~M.,  {Pavlenko} Y.~V.,   {Jones}
  H.~R.~A.,  2006, \mn@doi [\mnras] {10.1111/j.1365-2966.2005.09960.x}, \href
  {https://ui.adsabs.harvard.edu/abs/2006MNRAS.367..400H} {367, 400}

\bibitem[\protect\citeauthoryear{{Helling} \& {Casewell}}{{Helling} \&
  {Casewell}}{2014}]{Helling2014}
{Helling} C.,  {Casewell} S.,  2014, \mn@doi [\aapr]
  {10.1007/s00159-014-0080-0}, \href
  {https://ui.adsabs.harvard.edu/abs/2014A&ARv..22...80H} {22, 80}

\bibitem[\protect\citeauthoryear{{Helling}, {Woitke}  \& {Thi}}{{Helling}
  et~al.}{2008}]{Helling2008}
{Helling} C.,  {Woitke} P.,   {Thi} W.~F.,  2008, \mn@doi [\aap]
  {10.1051/0004-6361:20078220}, \href
  {https://ui.adsabs.harvard.edu/abs/2008A&A...485..547H} {485, 547}

\bibitem[\protect\citeauthoryear{{Heng}}{{Heng}}{2017}]{Heng2017}
{Heng} K.,  2017, {Exoplanetary Atmospheres: Theoretical Concepts and
  Foundations}.
Princeton University Press

\bibitem[\protect\citeauthoryear{{Hubeny} \& {Burrows}}{{Hubeny} \&
  {Burrows}}{2007}]{Hubeny2007}
{Hubeny} I.,  {Burrows} A.,  2007, \mn@doi [\apj] {10.1086/522107}, \href
  {https://ui.adsabs.harvard.edu/abs/2007ApJ...669.1248H} {669, 1248}

\bibitem[\protect\citeauthoryear{{Innes} \& {Pierrehumbert}}{{Innes} \&
  {Pierrehumbert}}{2022}]{Innes2022}
{Innes} H.,  {Pierrehumbert} R.~T.,  2022, \mn@doi [\apj]
  {10.3847/1538-4357/ac4887}, \href
  {https://ui.adsabs.harvard.edu/abs/2022ApJ...927...38I} {927, 38}

\bibitem[\protect\citeauthoryear{{Karalidi}, {Marley}, {Fortney}, {Morley},
  {Saumon}, {Lupu}, {Visscher}  \& {Freedman}}{{Karalidi}
  et~al.}{2021}]{Karalidi2021}
{Karalidi} T.,  {Marley} M.,  {Fortney} J.~J.,  {Morley} C.,  {Saumon} D.,
  {Lupu} R.,  {Visscher} C.,   {Freedman} R.,  2021, \mn@doi [\apj]
  {10.3847/1538-4357/ac3140}, \href
  {https://ui.adsabs.harvard.edu/abs/2021ApJ...923..269K} {923, 269}

\bibitem[\protect\citeauthoryear{{Karman} et~al.,}{{Karman}
  et~al.}{2019}]{Karman2019}
{Karman} T.,  et~al., 2019, \mn@doi [\icarus] {10.1016/j.icarus.2019.02.034},
  \href {https://ui.adsabs.harvard.edu/abs/2019Icar..328..160K} {328, 160}

\bibitem[\protect\citeauthoryear{{Komacek} \& {Showman}}{{Komacek} \&
  {Showman}}{2016}]{Komacek2016}
{Komacek} T.~D.,  {Showman} A.~P.,  2016, \mn@doi [\apj]
  {10.3847/0004-637X/821/1/16}, \href
  {https://ui.adsabs.harvard.edu/abs/2016ApJ...821...16K} {821, 16}

\bibitem[\protect\citeauthoryear{{Lacy} \& {Burrows}}{{Lacy} \&
  {Burrows}}{2023}]{Lacy2023}
{Lacy} B.,  {Burrows} A.,  2023, arXiv e-prints, \href
  {https://ui.adsabs.harvard.edu/abs/2023arXiv230316295L} {p. arXiv:2303.16295}

\bibitem[\protect\citeauthoryear{{Lee}, {Casewell}, {Chubb}, {Hammond}, {Tan},
  {Tsai}  \& {Pierrehumbert}}{{Lee} et~al.}{2020}]{Lee2020}
{Lee} E. K.~H.,  {Casewell} S.~L.,  {Chubb} K.~L.,  {Hammond} M.,  {Tan} X.,
  {Tsai} S.-M.,   {Pierrehumbert} R.~T.,  2020, \mn@doi [\mnras]
  {10.1093/mnras/staa1882}, \href
  {https://ui.adsabs.harvard.edu/abs/2020MNRAS.496.4674L} {496, 4674}

\bibitem[\protect\citeauthoryear{{Lee}, {Parmentier}, {Hammond}, {Grimm},
  {Kitzmann}, {Tan}, {Tsai}  \& {Pierrehumbert}}{{Lee} et~al.}{2021}]{Lee2021}
{Lee} E. K.~H.,  {Parmentier} V.,  {Hammond} M.,  {Grimm} S.~L.,  {Kitzmann}
  D.,  {Tan} X.,  {Tsai} S.-M.,   {Pierrehumbert} R.~T.,  2021, \mn@doi
  [\mnras] {10.1093/mnras/stab1851}, \href
  {https://ui.adsabs.harvard.edu/abs/2021MNRAS.506.2695L} {506, 2695}

\bibitem[\protect\citeauthoryear{{Lee} et~al.,}{{Lee} et~al.}{2022}]{Lee2022}
{Lee} E. K.~H.,  et~al., 2022, \mn@doi [\apj] {10.3847/1538-4357/ac61d6}, \href
  {https://ui.adsabs.harvard.edu/abs/2022ApJ...929..180L} {929, 180}

\bibitem[\protect\citeauthoryear{{Lee}, {Tsai}, {Hammond}  \& {Tan}}{{Lee}
  et~al.}{2023}]{Lee2023}
{Lee} E. K.~H.,  {Tsai} S.-M.,  {Hammond} M.,   {Tan} X.,  2023, \mn@doi [\aap]
  {10.1051/0004-6361/202245473}, \href
  {https://ui.adsabs.harvard.edu/abs/2023A&A...672A.110L} {672, A110}

\bibitem[\protect\citeauthoryear{{Lef{\`e}vre}, {Tan}, {Lee}  \&
  {Pierrehumbert}}{{Lef{\`e}vre} et~al.}{2022}]{Lefevre2022}
{Lef{\`e}vre} M.,  {Tan} X.,  {Lee} E. K.~H.,   {Pierrehumbert} R.~T.,  2022,
  \mn@doi [\apj] {10.3847/1538-4357/ac5e2d}, \href
  {https://ui.adsabs.harvard.edu/abs/2022ApJ...929..153L} {929, 153}

\bibitem[\protect\citeauthoryear{{Leggett} et~al.,}{{Leggett}
  et~al.}{2021}]{Leggett2021}
{Leggett} S.~K.,  et~al., 2021, \mn@doi [\apj] {10.3847/1538-4357/ac0cfe},
  \href {https://ui.adsabs.harvard.edu/abs/2021ApJ...918...11L} {918, 11}

\bibitem[\protect\citeauthoryear{{Li}, {Gordon}, {Rothman}, {Tan}, {Hu},
  {Kassi}, {Campargue}  \& {Medvedev}}{{Li} et~al.}{2015}]{Li2015}
{Li} G.,  {Gordon} I.~E.,  {Rothman} L.~S.,  {Tan} Y.,  {Hu} S.-M.,  {Kassi}
  S.,  {Campargue} A.,   {Medvedev} E.~S.,  2015, \mn@doi [The Astrophysical
  Journal Supplement Series] {10.1088/0067-0049/216/1/15}, \href
  {https://ui.adsabs.harvard.edu/#abs/2015ApJS..216...15L} {216, 15}

\bibitem[\protect\citeauthoryear{{Lian}, {Showman}, {Tan}  \& {Hu}}{{Lian}
  et~al.}{2022}]{lian2022}
{Lian} Y.,  {Showman} A.~P.,  {Tan} X.,   {Hu} Y.,  2022, \mn@doi [\apj]
  {10.3847/1538-4357/ac5598}, \href
  {https://ui.adsabs.harvard.edu/abs/2022ApJ...928..166L} {928, 166}

\bibitem[\protect\citeauthoryear{{Liu} \& {Showman}}{{Liu} \&
  {Showman}}{2013}]{Liu2013}
{Liu} B.,  {Showman} A.~P.,  2013, \mn@doi [\apj] {10.1088/0004-637X/770/1/42},
  \href {https://ui.adsabs.harvard.edu/abs/2013ApJ...770...42L} {770, 42}

\bibitem[\protect\citeauthoryear{{Mang}, {Gao}, {Hood}, {Fortney}, {Batalha},
  {Yu}  \& {de Pater}}{{Mang} et~al.}{2022}]{Mang2022}
{Mang} J.,  {Gao} P.,  {Hood} C.~E.,  {Fortney} J.~J.,  {Batalha} N.,  {Yu} X.,
    {de Pater} I.,  2022, \mn@doi [\apj] {10.3847/1538-4357/ac51d3}, \href
  {https://ui.adsabs.harvard.edu/abs/2022ApJ...927..184M} {927, 184}

\bibitem[\protect\citeauthoryear{{Manjavacas} et~al.,}{{Manjavacas}
  et~al.}{2019}]{Manjavacas2019}
{Manjavacas} E.,  et~al., 2019, \mn@doi [\apjl] {10.3847/2041-8213/ab13b9},
  \href {https://ui.adsabs.harvard.edu/abs/2019ApJ...875L..15M} {875, L15}

\bibitem[\protect\citeauthoryear{{Marley} \& {Robinson}}{{Marley} \&
  {Robinson}}{2015}]{Marley2015}
{Marley} M.~S.,  {Robinson} T.~D.,  2015, \mn@doi [\araa]
  {10.1146/annurev-astro-082214-122522}, \href
  {https://ui.adsabs.harvard.edu/abs/2015ARA&A..53..279M} {53, 279}

\bibitem[\protect\citeauthoryear{{Marley}, {Saumon}, {Guillot}, {Freedman},
  {Hubbard}, {Burrows}  \& {Lunine}}{{Marley} et~al.}{1996}]{Marley1996}
{Marley} M.~S.,  {Saumon} D.,  {Guillot} T.,  {Freedman} R.~S.,  {Hubbard}
  W.~B.,  {Burrows} A.,   {Lunine} J.~I.,  1996, \mn@doi [Science]
  {10.1126/science.272.5270.1919}, \href
  {https://ui.adsabs.harvard.edu/abs/1996Sci...272.1919M} {272, 1919}

\bibitem[\protect\citeauthoryear{{Marley}, {Seager}, {Saumon}, {Lodders},
  {Ackerman}, {Freedman}  \& {Fan}}{{Marley} et~al.}{2002}]{Marley2002}
{Marley} M.~S.,  {Seager} S.,  {Saumon} D.,  {Lodders} K.,  {Ackerman} A.~S.,
  {Freedman} R.~S.,   {Fan} X.,  2002, \mn@doi [\apj] {10.1086/338800}, \href
  {https://ui.adsabs.harvard.edu/abs/2002ApJ...568..335M} {568, 335}

\bibitem[\protect\citeauthoryear{{Marley} et~al.,}{{Marley}
  et~al.}{2021}]{Marley2021}
{Marley} M.~S.,  et~al., 2021, \mn@doi [\apj] {10.3847/1538-4357/ac141d}, \href
  {https://ui.adsabs.harvard.edu/abs/2021ApJ...920...85M} {920, 85}

\bibitem[\protect\citeauthoryear{{Mayne} et~al.,}{{Mayne}
  et~al.}{2014}]{Mayne2014}
{Mayne} N.~J.,  et~al., 2014, \mn@doi [\aap] {10.1051/0004-6361/201322174},
  \href {https://ui.adsabs.harvard.edu/abs/2014A&A...561A...1M} {561, A1}

\bibitem[\protect\citeauthoryear{{Metchev} et~al.,}{{Metchev}
  et~al.}{2015}]{Metchev2015}
{Metchev} S.~A.,  et~al., 2015, \mn@doi [\apj] {10.1088/0004-637X/799/2/154},
  \href {https://ui.adsabs.harvard.edu/abs/2015ApJ...799..154M} {799, 154}

\bibitem[\protect\citeauthoryear{{Miles} et~al.,}{{Miles}
  et~al.}{2020}]{Miles2020}
{Miles} B.~E.,  et~al., 2020, \mn@doi [\aj] {10.3847/1538-3881/ab9114}, \href
  {https://ui.adsabs.harvard.edu/abs/2020AJ....160...63M} {160, 63}

\bibitem[\protect\citeauthoryear{{Miles} et~al.,}{{Miles}
  et~al.}{2023}]{Miles2023}
{Miles} B.~E.,  et~al., 2023, \mn@doi [\apjl] {10.3847/2041-8213/acb04a}, \href
  {https://ui.adsabs.harvard.edu/abs/2023ApJ...946L...6M} {946, L6}

\bibitem[\protect\citeauthoryear{{Morley}, {Fortney}, {Marley}, {Visscher},
  {Saumon}  \& {Leggett}}{{Morley} et~al.}{2012}]{Morley2012}
{Morley} C.~V.,  {Fortney} J.~J.,  {Marley} M.~S.,  {Visscher} C.,  {Saumon}
  D.,   {Leggett} S.~K.,  2012, \mn@doi [\apj] {10.1088/0004-637X/756/2/172},
  \href {https://ui.adsabs.harvard.edu/abs/2012ApJ...756..172M} {756, 172}

\bibitem[\protect\citeauthoryear{{Morley}, {Marley}, {Fortney}, {Lupu},
  {Saumon}, {Greene}  \& {Lodders}}{{Morley} et~al.}{2014}]{Morley2014}
{Morley} C.~V.,  {Marley} M.~S.,  {Fortney} J.~J.,  {Lupu} R.,  {Saumon} D.,
  {Greene} T.,   {Lodders} K.,  2014, \mn@doi [\apj]
  {10.1088/0004-637X/787/1/78}, \href
  {https://ui.adsabs.harvard.edu/abs/2014ApJ...787...78M} {787, 78}

\bibitem[\protect\citeauthoryear{{Morley} et~al.,}{{Morley}
  et~al.}{2018}]{Morley2018}
{Morley} C.~V.,  et~al., 2018, \mn@doi [\apj] {10.3847/1538-4357/aabe8b}, \href
  {https://ui.adsabs.harvard.edu/abs/2018ApJ...858...97M} {858, 97}

\bibitem[\protect\citeauthoryear{{Mukherjee}, {Fortney}, {Batalha}, {Karalidi},
  {Marley}, {Visscher}, {Miles}  \& {Skemer}}{{Mukherjee}
  et~al.}{2022}]{Mukherjee2022}
{Mukherjee} S.,  {Fortney} J.~J.,  {Batalha} N.~E.,  {Karalidi} T.,  {Marley}
  M.~S.,  {Visscher} C.,  {Miles} B.~E.,   {Skemer} A. J.~I.,  2022, \mn@doi
  [\apj] {10.3847/1538-4357/ac8dfb}, \href
  {https://ui.adsabs.harvard.edu/abs/2022ApJ...938..107M} {938, 107}

\bibitem[\protect\citeauthoryear{{Noll}, {Geballe}  \& {Marley}}{{Noll}
  et~al.}{1997}]{Noll1997}
{Noll} K.~S.,  {Geballe} T.~R.,   {Marley} M.~S.,  1997, \mn@doi [\apjl]
  {10.1086/310954}, \href
  {https://ui.adsabs.harvard.edu/abs/1997ApJ...489L..87N} {489, L87}

\bibitem[\protect\citeauthoryear{{Oppenheimer}, {Kulkarni}, {Matthews}  \& {van
  Kerkwijk}}{{Oppenheimer} et~al.}{1998}]{Oppenheimer1998}
{Oppenheimer} B.~R.,  {Kulkarni} S.~R.,  {Matthews} K.,   {van Kerkwijk} M.~H.,
   1998, \mn@doi [\apj] {10.1086/305928}, \href
  {https://ui.adsabs.harvard.edu/abs/1998ApJ...502..932O} {502, 932}

\bibitem[\protect\citeauthoryear{{Phillips} et~al.,}{{Phillips}
  et~al.}{2020}]{Phillips2020}
{Phillips} M.~W.,  et~al., 2020, \mn@doi [\aap] {10.1051/0004-6361/201937381},
  \href {https://ui.adsabs.harvard.edu/abs/2020A&A...637A..38P} {637, A38}

\bibitem[\protect\citeauthoryear{{Polyansky}, {Kyuberis}, {Zobov}, {Tennyson},
  {Yurchenko}  \& {Lodi}}{{Polyansky} et~al.}{2018}]{Polyansky2018}
{Polyansky} O.~L.,  {Kyuberis} A.~A.,  {Zobov} N.~F.,  {Tennyson} J.,
  {Yurchenko} S.~N.,   {Lodi} L.,  2018, \mn@doi [\mnras]
  {10.1093/mnras/sty1877}, \href
  {http://adsabs.harvard.edu/abs/2018MNRAS.480.2597P} {480, 2597}

\bibitem[\protect\citeauthoryear{{Prinn} \& {Barshay}}{{Prinn} \&
  {Barshay}}{1977}]{Prinn1977}
{Prinn} R.~G.,  {Barshay} S.~S.,  1977, \mn@doi [Science]
  {10.1126/science.198.4321.1031}, \href
  {https://ui.adsabs.harvard.edu/abs/1977Sci...198.1031P} {198, 1031}

\bibitem[\protect\citeauthoryear{{Radigan}, {Lafreni{\`e}re}, {Jayawardhana}
  \& {Artigau}}{{Radigan} et~al.}{2014}]{Radigan2014}
{Radigan} J.,  {Lafreni{\`e}re} D.,  {Jayawardhana} R.,   {Artigau} E.,  2014,
  \mn@doi [\apj] {10.1088/0004-637X/793/2/75}, \href
  {https://ui.adsabs.harvard.edu/abs/2014ApJ...793...75R} {793, 75}

\bibitem[\protect\citeauthoryear{Reiners \& Basri}{Reiners \&
  Basri}{2008}]{reiners2008}
Reiners A.,  Basri G.,  2008, The Astrophysical Journal, 684, 1390

\bibitem[\protect\citeauthoryear{Rhines}{Rhines}{1975}]{rhines1975}
Rhines P.~B.,  1975, Journal of Fluid Mechanics, 69, 417

\bibitem[\protect\citeauthoryear{{Rustamkulov} et~al.,}{{Rustamkulov}
  et~al.}{2023}]{Rustamkulov2023}
{Rustamkulov} Z.,  et~al., 2023, \mn@doi [\nat] {10.1038/s41586-022-05677-y},
  \href {https://ui.adsabs.harvard.edu/abs/2023Natur.614..659R} {614, 659}

\bibitem[\protect\citeauthoryear{{Saumon}, {Geballe}, {Leggett}, {Marley},
  {Freedman}, {Lodders}, {Fegley}  \& {Sengupta}}{{Saumon}
  et~al.}{2000}]{Saumon2000}
{Saumon} D.,  {Geballe} T.~R.,  {Leggett} S.~K.,  {Marley} M.~S.,  {Freedman}
  R.~S.,  {Lodders} K.,  {Fegley} B. J.,   {Sengupta} S.~K.,  2000, \mn@doi
  [\apj] {10.1086/309410}, \href
  {https://ui.adsabs.harvard.edu/abs/2000ApJ...541..374S} {541, 374}

\bibitem[\protect\citeauthoryear{{Saumon}, {Marley}, {Abel}, {Frommhold}  \&
  {Freedman}}{{Saumon} et~al.}{2012}]{Saumon2012}
{Saumon} D.,  {Marley} M.~S.,  {Abel} M.,  {Frommhold} L.,   {Freedman} R.~S.,
  2012, \mn@doi [\apj] {10.1088/0004-637X/750/1/74}, \href
  {https://ui.adsabs.harvard.edu/abs/2012ApJ...750...74S} {750, 74}

\bibitem[\protect\citeauthoryear{{Showman}, {Tan}  \& {Zhang}}{{Showman}
  et~al.}{2019}]{Showman2019}
{Showman} A.~P.,  {Tan} X.,   {Zhang} X.,  2019, \mn@doi [\apj]
  {10.3847/1538-4357/ab384a}, \href
  {https://ui.adsabs.harvard.edu/abs/2019ApJ...883....4S} {883, 4}

\bibitem[\protect\citeauthoryear{{Stephens} et~al.,}{{Stephens}
  et~al.}{2009}]{Stephens2009}
{Stephens} D.~C.,  et~al., 2009, \mn@doi [\apj] {10.1088/0004-637X/702/1/154},
  \href {https://ui.adsabs.harvard.edu/abs/2009ApJ...702..154S} {702, 154}

\bibitem[\protect\citeauthoryear{{Tan}}{{Tan}}{2022}]{Tan2022}
{Tan} X.,  2022, \mn@doi [\mnras] {10.1093/mnras/stac344}, \href
  {https://ui.adsabs.harvard.edu/abs/2022MNRAS.511.4861T} {511, 4861}

\bibitem[\protect\citeauthoryear{{Tan} \& {Showman}}{{Tan} \&
  {Showman}}{2021a}]{Tan2021b}
{Tan} X.,  {Showman} A.~P.,  2021a, \mn@doi [\mnras] {10.1093/mnras/stab060},
  \href {https://ui.adsabs.harvard.edu/abs/2021MNRAS.502..678T} {502, 678}

\bibitem[\protect\citeauthoryear{{Tan} \& {Showman}}{{Tan} \&
  {Showman}}{2021b}]{Tan2021}
{Tan} X.,  {Showman} A.~P.,  2021b, \mn@doi [\mnras] {10.1093/mnras/stab097},
  \href {https://ui.adsabs.harvard.edu/abs/2021MNRAS.502.2198T} {502, 2198}

\bibitem[\protect\citeauthoryear{Tannock et~al.,}{Tannock
  et~al.}{2021}]{tannock2021}
Tannock M.~E.,  et~al., 2021, The Astronomical Journal, 161, 224

\bibitem[\protect\citeauthoryear{{Tsai}, {Lyons}, {Grosheintz}, {Rimmer},
  {Kitzmann}  \& {Heng}}{{Tsai} et~al.}{2017}]{Tsai2017}
{Tsai} S.-M.,  {Lyons} J.~R.,  {Grosheintz} L.,  {Rimmer} P.~B.,  {Kitzmann}
  D.,   {Heng} K.,  2017, \mn@doi [\apjs] {10.3847/1538-4365/228/2/20}, \href
  {https://ui.adsabs.harvard.edu/abs/2017ApJS..228...20T} {228, 20}

\bibitem[\protect\citeauthoryear{{Tsai}, {Malik}, {Kitzmann}, {Lyons},
  {Fateev}, {Lee}  \& {Heng}}{{Tsai} et~al.}{2021}]{Tsai2021}
{Tsai} S.-M.,  {Malik} M.,  {Kitzmann} D.,  {Lyons} J.~R.,  {Fateev} A.,  {Lee}
  E.,   {Heng} K.,  2021, \mn@doi [\apj] {10.3847/1538-4357/ac29bc}, \href
  {https://ui.adsabs.harvard.edu/abs/2021ApJ...923..264T} {923, 264}

\bibitem[\protect\citeauthoryear{{Tsai} et~al.,}{{Tsai}
  et~al.}{2022a}]{Tsai2022b}
{Tsai} S.-M.,  et~al., 2022a, \mn@doi [arXiv e-prints]
  {10.48550/arXiv.2211.10490}, \href
  {https://ui.adsabs.harvard.edu/abs/2022arXiv221110490T} {p. arXiv:2211.10490}

\bibitem[\protect\citeauthoryear{{Tsai}, {Lee}  \& {Pierrehumbert}}{{Tsai}
  et~al.}{2022b}]{Tsai2022}
{Tsai} S.-M.,  {Lee} E. K.~H.,   {Pierrehumbert} R.,  2022b, \mn@doi [\aap]
  {10.1051/0004-6361/202142816}, \href
  {https://ui.adsabs.harvard.edu/abs/2022A&A...664A..82T} {664, A82}

\bibitem[\protect\citeauthoryear{{Venot}, {Bounaceur}, {Dobrijevic},
  {H{\'e}brard}, {Cavali{\'e}}, {Tremblin}, {Drummond}  \& {Charnay}}{{Venot}
  et~al.}{2019}]{Venot2019}
{Venot} O.,  {Bounaceur} R.,  {Dobrijevic} M.,  {H{\'e}brard} E.,
  {Cavali{\'e}} T.,  {Tremblin} P.,  {Drummond} B.,   {Charnay} B.,  2019,
  \mn@doi [\aap] {10.1051/0004-6361/201834861}, \href
  {https://ui.adsabs.harvard.edu/abs/2019A&A...624A..58V} {624, A58}

\bibitem[\protect\citeauthoryear{{Visscher} \& {Moses}}{{Visscher} \&
  {Moses}}{2011}]{Visscher2011}
{Visscher} C.,  {Moses} J.~I.,  2011, \mn@doi [\apj]
  {10.1088/0004-637X/738/1/72}, \href
  {https://ui.adsabs.harvard.edu/abs/2011ApJ...738...72V} {738, 72}

\bibitem[\protect\citeauthoryear{{Vos}, {Allers}  \& {Biller}}{{Vos}
  et~al.}{2017}]{vos2017}
{Vos} J.~M.,  {Allers} K.~N.,   {Biller} B.~A.,  2017, \mn@doi [\apj]
  {10.3847/1538-4357/aa73cf}, \href
  {https://ui.adsabs.harvard.edu/abs/2017ApJ...842...78V} {842, 78}

\bibitem[\protect\citeauthoryear{{Vos}, {Faherty}, {Gagn{\'e}}, {Marley},
  {Metchev}, {Gizis}, {Rice}  \& {Cruz}}{{Vos} et~al.}{2022}]{vos2022}
{Vos} J.~M.,  {Faherty} J.~K.,  {Gagn{\'e}} J.,  {Marley} M.,  {Metchev} S.,
  {Gizis} J.,  {Rice} E.~L.,   {Cruz} K.,  2022, \mn@doi [\apj]
  {10.3847/1538-4357/ac4502}, \href
  {https://ui.adsabs.harvard.edu/abs/2022ApJ...924...68V} {924, 68}

\bibitem[\protect\citeauthoryear{{Witte}, {Helling}  \& {Hauschildt}}{{Witte}
  et~al.}{2009}]{Witte2009}
{Witte} S.,  {Helling} C.,   {Hauschildt} P.~H.,  2009, \mn@doi [\aap]
  {10.1051/0004-6361/200811501}, \href
  {https://ui.adsabs.harvard.edu/abs/2009A&A...506.1367W} {506, 1367}

\bibitem[\protect\citeauthoryear{{Yurchenko}, {Mellor}, {Freedman}  \&
  {Tennyson}}{{Yurchenko} et~al.}{2020}]{Yurchenko2020}
{Yurchenko} S.~N.,  {Mellor} T.~M.,  {Freedman} R.~S.,   {Tennyson} J.,  2020,
  \mn@doi [\mnras] {10.1093/mnras/staa1874}, \href
  {https://ui.adsabs.harvard.edu/abs/2020MNRAS.496.5282Y} {496, 5282}

\bibitem[\protect\citeauthoryear{{Zahnle} \& {Marley}}{{Zahnle} \&
  {Marley}}{2014}]{Zahnle2014}
{Zahnle} K.~J.,  {Marley} M.~S.,  2014, \mn@doi [\apj]
  {10.1088/0004-637X/797/1/41}, \href
  {https://ui.adsabs.harvard.edu/abs/2014ApJ...797...41Z} {797, 41}

\bibitem[\protect\citeauthoryear{{Zamyatina} et~al.,}{{Zamyatina}
  et~al.}{2023}]{Zamyatina2023}
{Zamyatina} M.,  et~al., 2023, \mn@doi [\mnras] {10.1093/mnras/stac3432}, \href
  {https://ui.adsabs.harvard.edu/abs/2023MNRAS.519.3129Z} {519, 3129}

\bibitem[\protect\citeauthoryear{{Zhang}}{{Zhang}}{2020}]{Zhang2020}
{Zhang} X.,  2020, \mn@doi [Research in Astronomy and Astrophysics]
  {10.1088/1674-4527/20/7/99}, \href
  {https://ui.adsabs.harvard.edu/abs/2020RAA....20...99Z} {20, 099}

\bibitem[\protect\citeauthoryear{{de la Cruz Rodr{\'\i}guez} \& {Piskunov}}{{de
  la Cruz Rodr{\'\i}guez} \& {Piskunov}}{2013}]{delaCruz2013}
{de la Cruz Rodr{\'\i}guez} J.,  {Piskunov} N.,  2013, \mn@doi [\apj]
  {10.1088/0004-637X/764/1/33}, \href
  {https://ui.adsabs.harvard.edu/abs/2013ApJ...764...33D} {764, 33}

\makeatother
\end{thebibliography}





\bsp	
\label{lastpage}
\end{document}